\documentclass[10pt,onecolumn]{article}
\usepackage[square,sort&compress]{natbib}
\usepackage[pdftex]{color}
\usepackage{graphicx}
\usepackage{amssymb}                  
\usepackage{txfonts}

\begin{document}

\def\aap{AAP }                                  
\def\aaps{AAPS }                                
\def\araa{ARA\&A }                              
\def\jrasc{JRASC }                              
\def\rvmp{RvMP }                                
\def\aar{Astron. Astrophys. Rev. }
\def\physrep{Phys. Rep. }                       
\def\ssr{Space Sci. Rev. }
\def\nat{Nature}
\def\mnras{MNRAS}
\def\apj{ApJ}
\def\apjl{ApJL}

\definecolor{purple1}{rgb}{0.4,0.,0.6}
\definecolor{purple2}{rgb}{0.6,0.,0.4}
\definecolor{brown}{rgb}{0.7,0.3,0.}
\definecolor{turquoise}{rgb}{0.,0.3,0.7}
\newcommand{\boldvec}[1]{\vec{\mbox{\boldmath{$#1$}}}}
\newcommand{\apropto}{\ \widetilde{\propto} \ }

\title{\bf Plasma turbulence in the interstellar medium}
\author{\underline{Katia Ferri\`{e}re$^1$} 
\medskip \\
$^1$ IRAP, Universit\'{e} de Toulouse, CNRS, Toulouse, France}
\date{Received 15 July 2019; accepted 1 October 2019}

\maketitle

\vspace*{-5.3truecm}
\leftline{Plasma Phys. Control. Fusion {\bf 62} (2020) 014014 (11pp)}
\vspace*{5truecm}

\begin{abstract}
The interstellar medium is a multi-phase, magnetized, and highly turbulent medium.
In this paper, we address both theoretical and observational aspects 
of plasma turbulence in the interstellar medium. 
We successively consider radio wave propagation through a plasma and radio polarized emission.
For each, we first provide a theoretical framework in the form of a few basic equations,
which enable us to define useful diagnostic tools of plasma turbulence;
we then show how these tools have been utilized to detect and interpret
observational signatures of plasma turbulence,
and what astronomers have learned from them regarding the nature, the sources,
and the dissipation of turbulence in the interstellar medium.
\end{abstract}

\section{Introduction}
\label{sect_intro}

The roughly 200 billion stars of our Galaxy are embedded in an extremely tenuous 
interstellar medium (ISM), which contains ordinary matter (made of gas and dust), 
cosmic rays, and magnetic fields.
These three basic constituents have comparable pressures,
and they are intimately coupled together by electromagnetic forces.
Through this coupling, magnetic fields affect both the spatial distribution
and the dynamics of the ordinary matter at all scales,
providing, in particular, efficient support against gravitational collapse
(see review by \cite{ferriere_01}).

Central to the life of the Galaxy is a cycle of matter and energy 
between the stars and the ISM. 
New stars continually form in the densest and coldest (molecular) regions of the ISM, 
where self-gravity becomes so strong that it overcomes magnetic support. 
Stars then undergo thermonuclear reactions, which enrich them in heavy elements.
A fraction of the enriched material eventually returns to the ISM 
through stellar winds and (mainly for the most massive stars) supernova explosions.
In both cases, the injection of stellar matter into the ISM is accompanied
by a strong release of energy, which generates turbulent motions in the ISM 
and maintains its heterogeneous structure.
To close the loop, some of the dense and cold regions produced by stellar feedback 
become the sites of a new generation of stars. 

Stellar feedback is believed to be the main source of interstellar turbulence,
in the sense that it dominates the total power input,
and the associated turbulent energy is presumably injected on scales
comparable to the radius of a typical wind bubble or supernova remnant, 
i.e., $\sim (10-100)~{\rm pc}$.
However, many other processes, operating over a much broader range of scales, 
also contribute \citep{elmegreen&s_04}.
At the largest ($\gtrsim 1~{\rm kpc}$) scales, Galactic rotation generates turbulence
at the shocks of spiral arms and bars as well as through various instabilities, 
including the shear and magneto-rotational instabilities.
At intermediate scales, aside from stellar feedback and still in the context 
of star formation, comes the gravitational instability.
At small ($\ll 1~{\rm pc}$) scales, cosmic-ray streaming drives plasma instabilities,
which lead to cosmic-ray acceleration and magnetic field amplification.

There exists ample observational evidence that the ISM is indeed turbulent
\citep{elmegreen&s_04,brandenburg&l_13}.
Historically, the first piece of evidence came from the finding that 
spectral lines from dense molecular regions have strongly superthermal widths
\citep{munch_58}.
Since then, high-resolution spectroscopic observations have been routinely used,
in combination with theoretical modeling, to recover the statistical properties 
of turbulent velocities
\citep{falgarone&p_90,falgarone&pw_91,miville&jfb_03,lazarian&p_00,lazarian&p_06}.
In parallel, high-resolution imaging of the emission produced by various 
interstellar tracers has provided complementary information on the statistics
of density fluctuations \citep{chepurnov&l_10,hennebelle&f_12,clark&pm_19}.
More recently, Zeeman measurements \citep{crutcher&whft_10,crutcher_17}
as well as polarization observations of synchrotron \citep{gaensler&etal_11}
and dust thermal emission \citep{planck19}
have started to complete the picture by adding information on the statistics
of magnetic field fluctuations.

In this paper, after introducing the relevant parameters of the turbulent ISM
(Section~\ref{sect_parameters}),
we tackle interstellar plasma turbulence from two different perspectives.
In Section~\ref{sect_propag}, we focus on the (collisionless) plasma aspects 
of the ISM and address three important effects of radio wave propagation 
through a plasma:
dispersion of pulsar signals, which gives access to the free-electron density, 
$n_{\rm e}$ (Section~\ref{sect_dispersion}),
interstellar scattering, which makes it possible to diagnose fluctuations 
in $n_{\rm e}$ down to the smallest scales (Section~\ref{sect_scattering}),
and Faraday rotation, which leads to
the magnetic field component parallel to the line of sight, 
$B_\parallel$ (Section~\ref{sect_faraday}).
This section pertains mainly to the warm ionized phase of the ISM,
which encloses most of the free electrons.
In Section~\ref{sect_polar}, we focus on the magnetic aspects of the ISM 
and discuss radio polarized emission, which carries information on
the magnetic field projected onto the plane of the sky, $\boldvec{B}_\perp$.
We first present existing studies of synchrotron emission 
(Section~\ref{sect_synchr}),
and then describe the novel technique of Faraday tomography 
(Section~\ref{sect_tomo}).
This section concerns the entire magnetized ISM.

\section{Physical parameters of the turbulent ISM}
\label{sect_parameters}

Interstellar matter is primarily composed of hydrogen,
but it also contains helium ($\approx 10~\%$ by number of hydrogen) 
and heavier elements, called ``metals" ($\approx 0.15~\%$ by number of hydrogen).
Virtually all the hydrogen, all the helium, and approximately half the metals
(i.e., $\approx 99~\%$ of the total interstellar mass) exist
in the form of gas; the other half of the metals ($\approx 1~\%$ of
the interstellar mass) is locked up in small solid grains of dust.

Interstellar gas can be found in molecular, (cold and warm) atomic,
and (warm and hot) ionized forms.
The molecular gas constitutes the coldest and densest medium,
generally referred to as the molecular medium (MM).
The mostly neutral atomic gas exists in two distinct phases:
the cold neutral medium (CNM) and the warm neutral medium (WNM).
Likewise, the ionized gas is divided into a warm ionized medium (WIM) 
and a hot ionized medium (HIM).
Typical ranges of values for the temperature, $T$, hydrogen density, $n_{\rm H}$, 
and electron-to-hydrogen density ratio, 
$\frac{\displaystyle n_{\rm e}}{\displaystyle n_{\rm H}}$,
of these different phases are given in Table~\ref{table_ism_parameters}
\citep{ferriere_01}.

\begin{table}
\begin{tabular}{lccccc}
\hline
\noalign{\medskip}
& MM & CNM & WNM & WIM & HIM \\
\noalign{\medskip}
\hline
\noalign{\medskip}
$T~[{\rm K}]$ & 
$10 - 20$ & $50 - 100$ & $10^3 - 10^4$ & $\sim 10^4$ & $\sim 10^6$ \\
\noalign{\smallskip}
$n_{\rm H}~[{\rm cm}^{-3}]$ & 
$10^2 - 10^6$ & $20 - 50$ & $0.2 - 2$ & $0.1 - 0.5$ & $0.003 - 0.01$ \\
\noalign{\smallskip}
$\frac{\displaystyle n_{\rm e}}{\displaystyle n_{\rm H}}$ & 
$\lll 1$ & 
$(0.3 - 1) \, 10^{-3}$ & $0.01 - 0.05$ & $\approx 1$ & $\approx 1.2$ \\
\noalign{\medskip}
\hline
\end{tabular}
\caption{Basic physical parameters of the different ISM phases:
$T$ is the temperature, $n_{\rm H}$ the hydrogen density,
and $n_{\rm e}$ the free-electron density.
}
\label{table_ism_parameters}
\end{table}

In the following, we will not discuss the MM,
whose ionization fraction is extremely low.
The atomic phases, too, are weakly ionized, 
but their low ionization fraction is often sufficient 
for the neutrals to remain tightly coupled to the ions
through ion-neutral collisions.
For reference, the collision time of a neutral by an ion is typically 
$\sim (10^{2} - 10^{3})~{\rm yr}$ in the CNM 
and $\sim (10^{2} - 10^{4})~{\rm yr}$ in the WNM, 
both of which are very short by Galactic standards.

Despite the stark temperature and density contrasts between the different 
ISM phases, thermal pressure, $P_{\rm th}$, remains roughly uniform 
(within an order of magnitude) across them. 
Magnetic pressure, $P_{\rm m}$, is also roughly uniform throughout the ISM.
In this paper, we adopt the canonical value of the interstellar 
magnetic field strength, $B \simeq 5~\mu{\rm G}$, 
corresponding to $P_{\rm m} \simeq 10^{-12}~{\rm ergs~cm}^{-3}$.
It then follows that $P_{\rm th} \sim P_{\rm m}$, or equivalently, 
$\beta \equiv \frac{\displaystyle P_{\rm th}}{\displaystyle P_{\rm m}} \sim 1$.

In Table~\ref{table_plasma_parameters}, we choose a particular set of values 
for the temperature and densities of the different ISM phases, 
and we provide the associated values of a few key parameters directly relevant 
to interstellar turbulence, namely,
the isothermal sound speed, 
$C_{\rm s} = \sqrt{\frac{\displaystyle P_{\rm th}}{\displaystyle \rho}}$, 
the Alfv{\'e}n speed, 
$V_{\rm A} = \sqrt{\frac{\displaystyle 2 \, P_{\rm m}}{\displaystyle \rho}}$,
the plasma frequency, $\omega_{\rm e} = 
\sqrt{\frac{\displaystyle 4\pi n_{\rm e} e^2}{\displaystyle m_{\rm e}}}$, 
the electron gyro-frequency, $\Omega_{\rm e} = 
\frac{\displaystyle -e B}{\displaystyle m_{\rm e} c}$,
gyro-radius, $r_{\rm e} = 
\frac{\displaystyle \upsilon_{\rm \perp e}}{\displaystyle |\Omega_{\rm e}|}$,
collision time, $\tau_{\rm e} \apropto 
\frac{\displaystyle T^{1.5}}{\displaystyle n_{\rm e}}$, 
and collision mean free-path, $\lambda_{\rm e} \apropto 
\frac{\displaystyle T^2}{\displaystyle n_{\rm e}}$.
The collision parameters appear to vary widely from one phase to the next,
as expected from their inverse dependencies on temperature and electron density, 
together with the rough inverse proportionality between $T$ and $n_{\rm e}$.
In all cases, $\omega_{\rm e} \gg |\Omega_{\rm e}| 
\gg \tau_{\rm e}^{-1}$ 
and $r_{\rm e} \ll \lambda_{\rm e}$.

\begin{table}
\begin{tabular}{lccccc}
\hline
\noalign{\medskip}
& CNM & WNM & WIM & HIM \\
\noalign{\medskip}
\hline
\noalign{\medskip}
$T~[{\rm K}]$ & 
$80$ & $5\,000$ & $8\,000$ & $10^6$ \\
\noalign{\smallskip}
$n_{\rm H}~[{\rm cm}^{-3}]$ & 
$30$ & $0.4$ & $0.2$ & $0.005$ \\
\noalign{\smallskip}
$n_{\rm e}~[{\rm cm}^{-3}]$ &
$0.02$ & $0.01$ & $0.2$ & $0.006$ \\
\noalign{\medskip}
\hline
\noalign{\medskip}
$C_{\rm s}~[{\rm km~s^{-1}}]$ & 0.7 & 6 & 10 & 120 \\
\noalign{\smallskip}
$V_{\rm A}~[{\rm km~s^{-1}}]$ & 1.7 & 15 & 20 & 130 \\
\noalign{\smallskip}
\hline
\noalign{\medskip}
$\omega_{\rm e}~[{\rm kHz}]$ & 8 & 5.5 & 25 & 4 \\
\noalign{\medskip}
$|\Omega_{\rm e}|~[{\rm Hz}]$ & 90 & 90 & 90 & 90 \\
\noalign{\smallskip}
$r_{\rm e}~[{\rm km}]$ & 0.6 & 5 & 6 & 60 \\
\noalign{\medskip}
$\tau_{\rm e}$ & 9~{\rm min} & 5~{\rm d} & 12~{\rm hr} & 45~{\rm yr} \\
\noalign{\smallskip}
$\lambda_{\rm e}$ & $5~R_\oplus$ & $1.3~{\rm AU}$ & $0.17~{\rm AU}$ & 
$0.3~{\rm pc}$ \\
\noalign{\medskip}
\hline
\noalign{\medskip}
$L~[{\rm pc}]$ & 10 & 30 & 30 & 100 \\
\noalign{\smallskip}
$V~[{\rm km~s^{-1}}]$ & 3 & 10 & 10 & 30 \\
\noalign{\medskip}
\hline
\noalign{\medskip}
${\rm Re}$ & $4 \times 10^{10}$ & $10^{7}$ & $5 \times 10^{7}$ & $10^{2}$ \\
\noalign{\smallskip}
${\rm Re_m}$ & $7 \times 10^{14}$ & $3 \times 10^{18}$ & $5 \times 10^{18}$ & 
$5 \times 10^{22}$ \\
\noalign{\smallskip}
${\rm P_m}$ & $2 \times 10^{4}$ & $3 \times 10^{11}$ & $10^{11}$ & 
$5 \times 10^{20}$ \\
\noalign{\medskip}
\hline
\end{tabular}
\caption{Set of representative values for a few key parameters 
in the different ISM phases:
$T$ is the temperature, 
$n_{\rm H}$ the hydrogen density,
$n_{\rm e}$ the free-electron density, 
$C_{\rm s}$ the isothermal sound speed,
$V_{\rm A}$ the Alfv{\'e}n speed,
$\omega_{\rm e}$ the plasma frequency, 
$\Omega_{\rm e}$ the electron gyro-frequency,
$r_{\rm e}$ the electron gyro-radius,
$\tau_{\rm e}$ the electron collision time,
$\lambda_{\rm e}$ the electron collision mean free-path,
$L$ a typical length scale, $V$ a characteristic velocity,
${\rm Re}$ the standard (fluid) Reynolds number,
${\rm Re_m}$ its magnetic counterpart,
and ${\rm P_m}$ the magnetic Prandtl number.
The values of $V_{\rm A}$, $\Omega_{\rm e}$, and $r_{\rm e}$ 
are obtained for $B = 5~\mu{\rm G}$.
}
\label{table_plasma_parameters}
\end{table}

As will become apparent in Section~\ref{sect_scattering}, 
turbulent fluctuations in the ISM span a huge range of scales on either side of 
the proton collision mean free-path, $\lambda_{\rm p} \simeq \sqrt{2} \, \lambda_{\rm e}$.
Fluctuations with scales $\lesssim \lambda_{\rm p}$ require a plasma description,
while fluctuations with scales $\gg \lambda_{\rm p}$ can be studied with 
magnetohydrodynamics (MHD).

Also shown in Table~\ref{table_plasma_parameters} are the values 
of the standard (fluid) and magnetic Reynolds numbers,
${\rm Re} \equiv \frac{\displaystyle V \, L}{\displaystyle \nu}$ 
and ${\rm Re_m} \equiv \frac{\displaystyle V \, L}{\displaystyle \eta}$,
as well as the magnetic Prandtl number,
${\rm P_m} \equiv \frac{\displaystyle {\rm Re_m}}{\displaystyle {\rm Re}}$,
where $L$ is a typical length scale, $V$ a characteristic velocity, 
$\nu$ the kinematic viscosity, and $\eta$ the magnetic resistivity.
Like for the collision parameters, the enormous variations in the Reynolds numbers
and the even more dramatic variations in the magnetic Prandtl number
(16 orders of magnitude between the CNM and the HIM) 
can be explained by their dependencies on temperature and electron density:
${\rm Re} \propto \frac{\displaystyle 1}{\displaystyle \nu} 
\apropto \frac{\displaystyle n_{\rm e}}{\displaystyle T^{2.5}}$,
${\rm Re_m} \propto \frac{\displaystyle 1}{\displaystyle \eta} \apropto T^{1.5}$,
and ${\rm P_m}  = \frac{\displaystyle \nu}{\displaystyle \eta} 
\apropto \frac{\displaystyle T^4}{\displaystyle n_{\rm e}}$.
Besides, the huge values of the Reynolds numbers in the four different phases 
indicate that the entire ISM is in a state of fully developed turbulence
and that this turbulence can lead to magnetic field amplification (dynamo action). 
The Prandtl number has little impact at large scales, 
but in MHD fluids it governs the behavior of the turbulent cascade 
at the small dissipative scales:
large values of ${\rm P_m}$ imply that the kinetic-energy cascade 
is truncated by viscosity at scales much larger than the scales 
at which the magnetic-energy cascade is truncated by resistivity.
This situation has important implications for dynamo action in our Galaxy.

It also emerges from Table~\ref{table_plasma_parameters} 
that the sonic and Alfv{\'e}nic Mach numbers,
${\cal M}_{\rm s} \equiv \frac{\displaystyle V}{\displaystyle C_{\rm s}}$
and ${\cal M}_{\rm A} \equiv \frac{\displaystyle V}{\displaystyle V_{\rm A}}$,
are roughly $\sim 1$, which in turn indicates that turbulence is generally 
trans-sonic and trans-Alfv{\'e}nic. 
Only the HIM appears to be (slighty) sub-sonic and sub-Alfv{\'e}nic.

Altogether, the three dimensionless parameters $\beta$, ${\cal M}_{\rm s}$, 
and ${\cal M}_{\rm A}$ are all of order unity, which means that 
thermal, magnetic, and kinetic pressures (or energies) are all 
in rough equipartition. 

In Section~\ref{sect_propag}, we specifically focus on the WIM, 
which contains most of the free electrons and is, therefore,
the best-diagnosed ISM phase from a plasma perspective.
In Section~\ref{sect_polar}, we consider the magnetized ISM as a whole,
including its ionized and neutral (molecular and atomic) phases.
We do not discuss the latter separately, noting that they have generally 
been studied more as MHD fluids than for their plasma properties.

\section{Effects of radio wave propagation}
\label{sect_propag}

When a radio wave propagates through a plasma, say, the WIM,
it interacts with the free electrons of the plasma.
These interactions slow down its propagation, to an extent that depends 
on wavelength and on polarization direction.
This leads to a number of observable effects, 
such as temporal dispersion, scintillation, Faraday rotation...
Measuring these effects makes it possible to trace back 
to some properties of the traversed plasma.

Mathematically, the propagation of a radio electromagnetic wave 
parallel to the magnetic field, $\boldvec{B}$,
can be described by the dispersion relation,
\begin{equation}
\omega^2 \ = \ c^2 \, k^2 
\ + \ \frac{\omega_{\rm e}^2}
{1 \pm \frac{\displaystyle \Omega_{\rm e}}{\displaystyle\omega}} \ ,
\label{eq_DR}
\end{equation}
where $\omega$ and $k$ are the angular frequency and wavenumber of the wave, 
$c$ is the speed of light, 
$\omega_{\rm e} = 
\sqrt{\frac{\displaystyle 4\pi n_{\rm e} e^2}{\displaystyle m_{\rm e}}}$
the plasma frequency,
and $\Omega_{\rm e} = 
\frac{\displaystyle -e B}{\displaystyle m_{\rm e} c}$
the electron gyro-frequency.
The dispersion relation can also be written
in terms of the refractive index, ${\sf n}$:
\begin{equation}
{\sf n}^2 
\ \equiv \ \frac{c^2 \, k^2}{\omega^2} 
\ = \ 1 
\ - \ \frac{\frac{\displaystyle \omega_{\rm e}^2}{\displaystyle \omega^2}}
{1 \pm \frac{\displaystyle \Omega_{\rm e}}{\displaystyle \omega}} \ \cdot
\label{eq_refr_index}
\end{equation}
The waves of interest here are typically in the GHz regime, 
so that $|\Omega_{\rm e}| \ll \omega_{\rm e} \lll \omega$ 
(see Table~\ref{table_plasma_parameters}). 
Clearly, the first term on the right-hand sides of Eqs.~(\ref{eq_DR}) 
and (\ref{eq_refr_index}) describes wave propagation in vacuum.
The second term corrects for the presence of free electrons,
with a first-order correction 
$\propto \frac{\displaystyle \omega_{\rm e}^2}{\displaystyle \omega^2} 
\propto n_{\rm e}$
and a finer correction
$\propto 
\frac{\displaystyle |\Omega_{\rm e}|}{\displaystyle \omega} 
\propto B$.
It then follows that radio wave propagation is sensitive
primarily to the electron density, $n_{\rm e}$, 
and only secondarily to the magnetic field strength, $B$.
This is a little unfortunate given that density fluctuations provide only 
a rather indirect tracer of turbulence,
as opposed to velocity and magnetic field fluctuations
\citep{brandenburg&l_13}.

In Sections~\ref{sect_dispersion} and \ref{sect_scattering}, 
we examine the effects of the electron density and its fluctuations 
on radio wave propagation,
and in Section~\ref{sect_faraday}, we turn to the effects of the magnetic field.

\subsection{Dispersion of pulsar signals}
\label{sect_dispersion}

To first order in the plasma correction, the dispersion relation, 
Eq.~(\ref{eq_DR}), reduces to 
\begin{equation}
\omega^2 
\ = \ c^2 \, k^2 + \omega_{\rm e}^2
\ = \ c^2 \, k^2 + \frac{4\pi \, n_{\rm e} \, e^2}{m_{\rm e}} \ \cdot
\label{eq_DR_1st}
\end{equation}
The group velocity 
is then given by
\begin{equation}
V_{\rm g} 
\ = \ \frac{\partial \omega}{\partial k}
\ = \ c \ 
\left( 1 - \frac{\omega_{\rm e}^2}{2 \, \omega^2} \right)
\ = \ c \ \left( 1 - 
\frac{e^2}{2\pi \, m_{\rm e} \, c^2} \ n_{\rm e} \ \lambda^2
\right) \ ,
\label{eq_group_vel}
\end{equation}
which shows that the slowdown of wave propagation caused by the electrons 
increases linearly with electron density, $n_{\rm e}$,
and quadratically with wavelength, $\lambda$.
If we now consider a source of radio waves, 
the travel time from the source (src) to the observer (obs)
can be written as
\begin{equation}
t_{\rm tr}
\ = \ \int_{\rm src}^{\rm obs} \frac{ds}{V_{\rm g}}
\ = \ \frac{L}{c} 
\ + \ \frac{e^2}{2\pi \, m_{\rm e} \, c^3} \ 
{\rm DM} \
\lambda^2 \ ,
\label{eq_travel_time}
\end{equation}
where $L$ is the path length from the source to the observer and
\begin{equation}
{\rm DM} \ \equiv \ \int_{\rm src}^{\rm obs} n_{\rm e} \ ds
\label{eq_DM}
\end{equation}
is the so-called dispersion measure.

The most useful radio sources in the present context are Galactic pulsars.
A pulsar is a rapidly spinning, strongly magnetized neutron star, 
which appears to emit periodic pulses of radiation.
These pulses can each be decomposed into a spectrum of electromagnetic waves 
spanning a whole range of radio wavelengths.
As we just saw, the longer-wavelength waves propagate less rapidly
through interstellar space and, therefore, arrive slightly later at the observer.
By measuring the resulting spread in arrival times over a wavelength-squared range 
$\Delta \lambda^2$,
\begin{equation}
\Delta t_{\rm tr}
\ = \ \frac{e^2}{2\pi \, m_{\rm e} \, c^3} \ 
{\rm DM} \
\Delta \lambda^2 \ ,
\label{eq_spread_time}
\end{equation}
one can directly infer the pulsar dispersion measure, i.e., 
the column density of free electrons between the pulsar and the observer.

Since the discovery of the first pulsar \citep{hewish&bps_68},
astronomers have used pulsars with measured distances and dispersion measures 
to map out the large-scale distribution of free electrons in our Galaxy
\citep{cordes&l_02,schnitzeler_12,yao&mw_17}.
There are currently 2\,702 known pulsars, as listed in the latest version 
(1.60) of the ATNF pulsar catalogue.\footnote{
http://www.atnf.csiro.au/research/pulsar/psrcat
\citep{manchester&hth_05}. 
} 
Of these pulsars, 2\,607 have measured DMs and 1\,159 have measured RMs 
(defined later, in Section~\ref{sect_faraday}).
The Square Kilometer Array in phase~1 (SKA\,1) is expected to detect 
virtually all the Galactic pulsars pointing toward the Earth,
thereby bringing the number of known pulsars to $\approx 18\,000$, 
with a surface density toward the Galactic plane $\sim 6~{\rm deg}^{-2}$ 
\citep{keane&etal_15}.
Most of these pulsars will have measured DMs and RMs 
(Charlotte Sobey, private communication).
Even so, it is clear that the number of observable pulsars 
will never be sufficient for us to measure small-scale details 
in the electron density distribution.

The best avenue to obtain information on small-scale density fluctuations
is to consider the effects of interstellar scattering.

\subsection{Interstellar scattering}
\label{sect_scattering}

When a radio wave propagates through a {\it turbulent} plasma, 
it encounters stochastic density fluctuations, 
which induce fluctuations in the refractive index, 
which in turn cause phase modulations.
As a result, wavefronts become randomly distorted --
in other words, the wave gets scattered.
The scattered waves then interfere alternatively constructively and destructively,
thereby causing random fluctuations in the amplitude and phase 
of the received wave field.
These fluctuations, which are collectively known as interstellar scintillations (ISS),
provide very useful diagnostics of the turbulent properties of the traversed plasma
(see \cite{rickett_90} for a detailed review).

To first order in the plasma correction, the expression of the refractive index,
Eq.~(\ref{eq_refr_index}), reduces to
\begin{equation}
{\sf n}^2 
\ = \ 1 - \frac{\omega_{\rm e}^2}{\omega^2}
\ = \ 1 - \frac{4\pi \, n_{\rm e} \, e^2}{m_{\rm e} \ \omega^2} \ \cdot
\label{eq_refr_index_1st}
\end{equation}
Fluctuations in the refractive index, $\delta {\sf n}$, 
entail fluctuations in the phase $\phi$ of the received wave field,
\begin{equation}
\delta \phi
\ = \ \int_{\rm src}^{\rm obs} \delta {\sf n} \  k \ ds
\ = \ r_{\rm e} \ \lambda \
\left( \int_{\rm src}^{\rm obs} \delta n_{\rm e} \ ds \right) \ ,
\label{eq_phase_fluct}
\end{equation}
where $r_{\rm e} = \frac{\displaystyle e^2}{\displaystyle m_{\rm e} \, c^2}$
is the classical electron radius \citep{rickett_90}.

\begin{figure}
{\large (a)} 
\raisebox{-0.18\textheight}{
\includegraphics[width=0.4\textwidth]{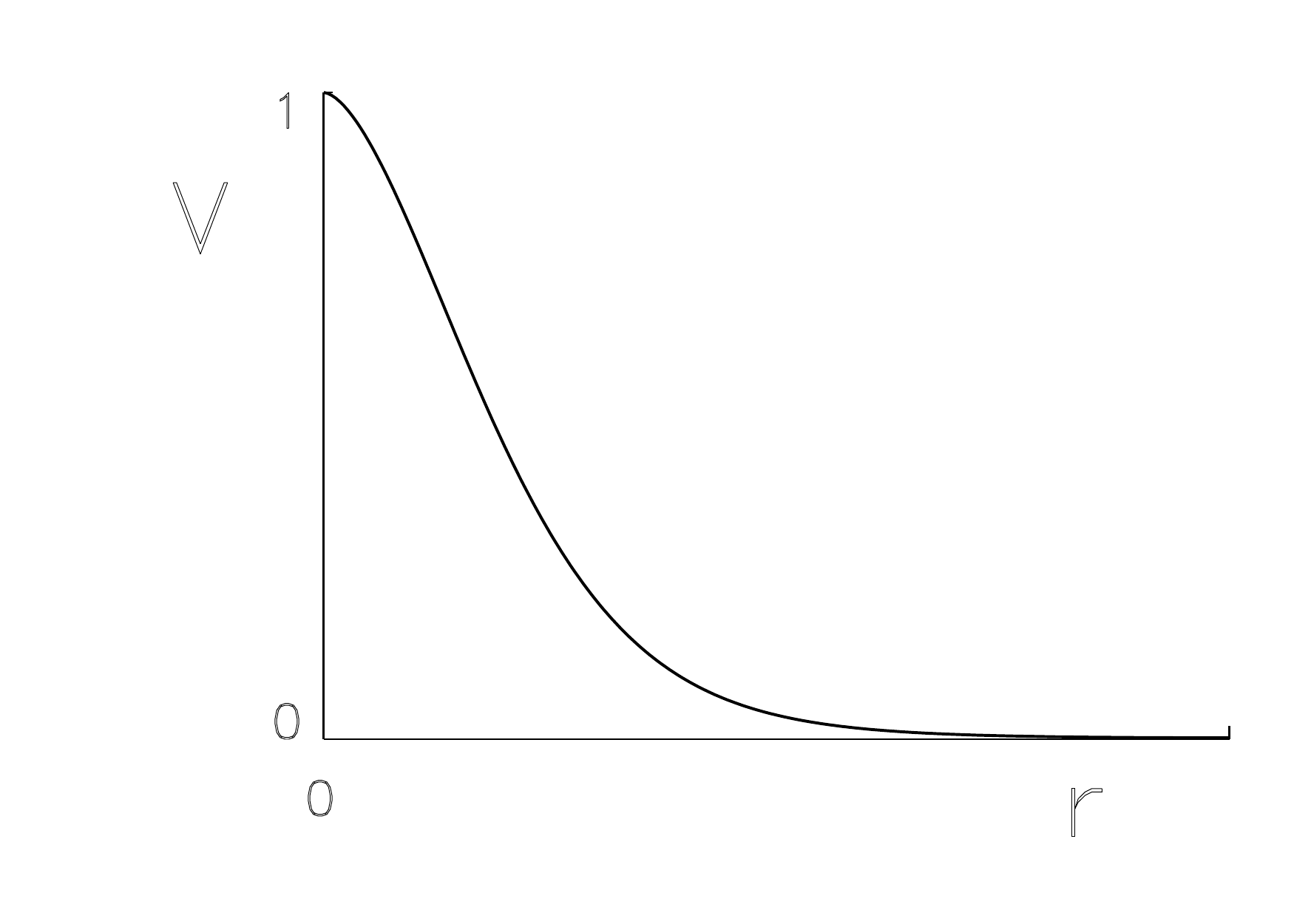}
}
\qquad
{\large (b)} 
\raisebox{-0.18\textheight}{
\includegraphics[width=0.4\textwidth]{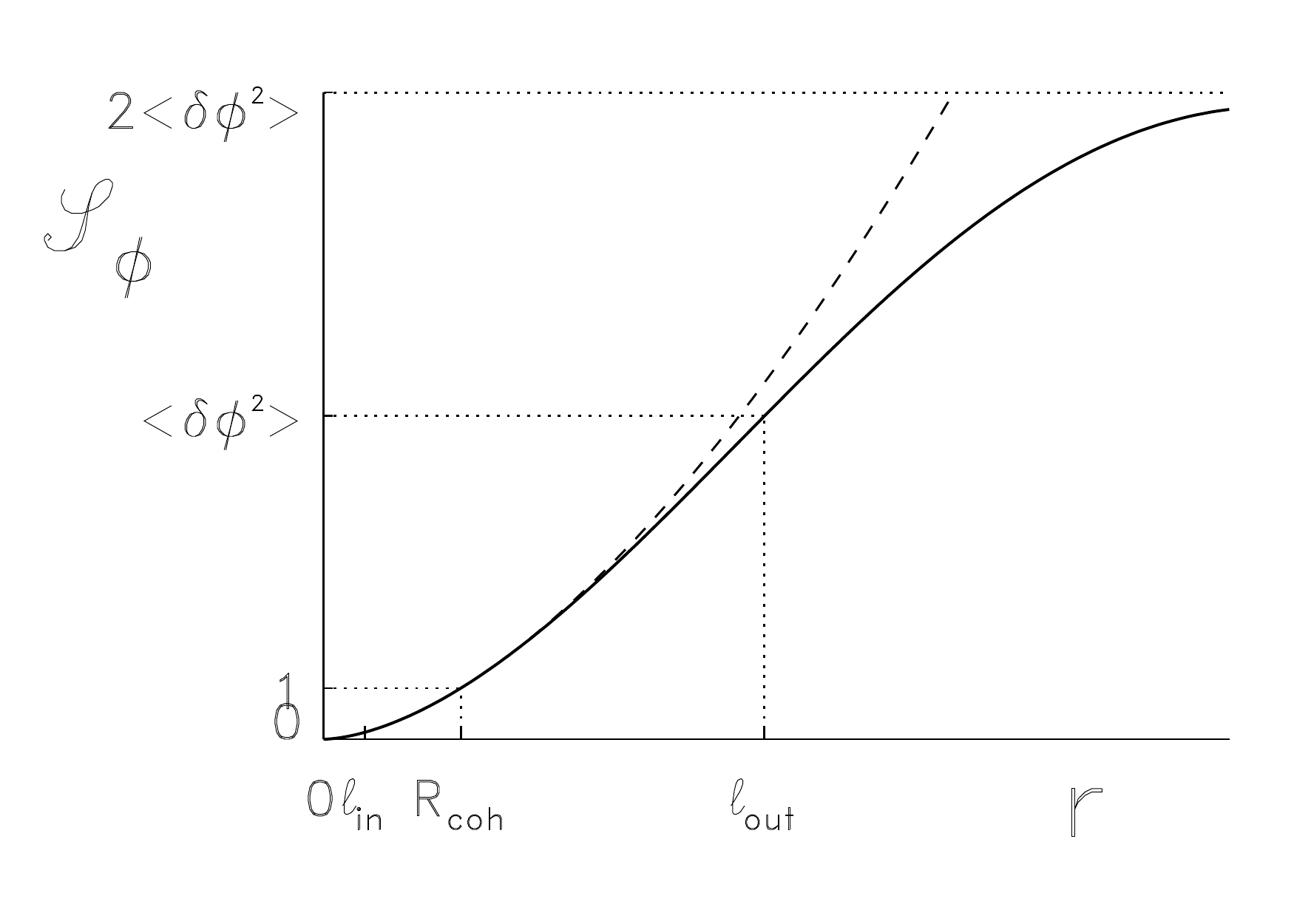}
}
\caption{(a) Visibility function, $V$, 
and (b) phase structure function, ${\cal S}_\phi$,
of a scattered radio source, 
both as functions of interferometer baseline length, $r$.
}
\label{figure_visibility}
\end{figure}

A convenient tool to study the statistics of phase fluctuations is
the phase structure function,
\begin{equation}
{\cal S}_\phi (\boldvec{r}) \ = \ 
{\bigg \langle} \Big[ 
\delta \phi (\boldvec{r}_1) - \delta \phi (\boldvec{r_2})
\Big]^2 {\bigg \rangle} \ ,
\label{eq_struct_fc}
\end{equation}
where $\boldvec{r}_1$ and $\boldvec{r}_2$ are the position vectors
(in a plane transverse to the line of sight)
of two points at which the wave field is being measured,
i.e., two antennas of a radio interferometer; 
$\boldvec{r} = \boldvec{r}_2 - \boldvec{r}_1$ 
is the separation vector between the two antennas,
i.e., the interferometer baseline vector (transverse to the line of sight);
and the angle brackets denote an ensemble average 
or, in practice, a temporal average over a sufficiently long time.
The phase structure function is a quantity that can be measured almost directly.
In fact, what a radio interferometer measures is the so-called 
complex visibility function,\footnote{
The complex visibility function of a scattered radio source, $V(u,\upsilon)$,
is the 2D Fourier transform of its brightness (or intensity) 
angular distribution in the sky, $I(\alpha,\delta)$.
A radio interferometer measures $V(u,\upsilon)$ from 
the correlation between the wave electric field at two antennas 
separated by $\boldvec{r}$ \citep{thompson&ms_86}.
}
$V(u,\upsilon)$, 
with $(u,\upsilon) = \frac{\displaystyle \boldvec{r}}{\displaystyle \lambda}$,
and this visibility function is directly related to the phase structure function 
through
\begin{equation}
V(\boldvec{r}) \ = \ 
\exp \left( - \textstyle{\frac{1}{2}} \, {\cal S}_\phi(\boldvec{r}) \right) 
\label{eq_visibility}
\end{equation}
\citep{cordes&wb_85}.
In the case of isotropic turbulence/scattering, 
both the visibility function, $V(\boldvec{r})$,
and the phase structure function, ${\cal S}_\phi(\boldvec{r})$, 
depend only on the distance between the two antennas, 
i.e., the interferometer baseline length, $r$.
For illustration, Figure~\ref{figure_visibility} displays typical profiles of $V(r)$ 
and ${\cal S}_\phi(r)$.
The important point to notice in Figure~\ref{figure_visibility} is that 
$V(r)$ decreases from $1$ at $r = 0$ to $0$ for $r \to \infty$, 
while ${\cal S}_\phi(r)$ increases from $0$ at $r = 0$ 
to twice the phase variance, $\langle \delta \phi ^2 \rangle$, 
for $r \to \infty$.

Once the phase structure function, ${\cal S}_\phi(\boldvec{r})$, 
has been measured, it can be used to extract information 
on the electron density fluctuations.
In general, the statistics of density fluctuations are described 
as a function of the fluctuation wavevector, $\boldvec{q}$,\footnote{
Throughout this paper, $\boldvec{k}$ denotes the wavevector of the incoming 
radio wave and $\boldvec{q}$ the wavevector of density (or magnetic field)
fluctuations in the traversed interstellar plasma. 
}
by the spatial power spectrum, $P_n(\boldvec{q})$,
defined such that
\begin{equation}
{\big \langle} \delta n_{\rm e}^2 {\big \rangle}
\ = \ \int P_n(\boldvec{q}) \ d\boldvec{q} \ \cdot
\label{eq_variance}
\end{equation}
In the case of isotropic turbulence, $P_n(\boldvec{q}) \, d\boldvec{q} 
= 4\pi \, q^2 \, P_n(q) \, dq$, 
and the 3D power spectrum, $P_n(q)$, can be replaced by
the 1D power spectrum, $E_n(q) \equiv 4\pi \, q^2 \, P_n(q)$,
so that Eq.~(\ref{eq_variance}) becomes
\begin{equation}
{\big \langle} \delta n_{\rm e}^2 {\big \rangle}
\ = \ \int_{0}^{\infty} E_n(q) \ dq \ \cdot
\label{eq_variance_isotr}
\end{equation}
It then follows from Eqs.~(\ref{eq_struct_fc}), (\ref{eq_phase_fluct}), 
and (\ref{eq_variance_isotr}) that
the phase structure function, ${\cal S}_\phi(r)$,
can be related to the electron density power spectrum, $E_n(q)$, through
\begin{equation}
{\cal S}_\phi(r) \ = \ 
r_{\rm e}^2 \ \lambda^2 \ 
\int_{\rm src}^{\rm obs} 
\left( \int_{0}^{\infty} F(q r) \ \frac{E_n(q)}{q} \ dq \right) 
\ ds \ ,
\label{eq_struct_fc_isotr}
\end{equation}
where $F(q r) = 2\pi \, [1 - J_0(q r)]$
\citep{cordes&wb_85}.

Under typical interstellar conditions ($E_n(q) \propto q^{-5/3}$; see below), 
the integral over $q$ in Eq.~(\ref{eq_struct_fc_isotr}) 
is dominated by contributions from $q r \sim 1$. 
Physically, the phase structure function measured 
at a given baseline length $r$ is mostly sensitive to, and hence mainly probes, 
density fluctuations with scales $\frac{\displaystyle 1}{\displaystyle q} \sim r$
\citep{haverkorn&s_13}.
This automatically sets an upper limit $\sim 10^{4}~{\rm km}$ 
(reached with {\it Very Long Baseline Interferometry} (VLBI) arrays) 
to the scales of density fluctuations that can be measured with ${\cal S}_\phi$.
Although large by terrestrial standards, these maximum scales remain tiny 
for the ISM.

The electron density power spectrum can often be described by a power law of $q$:
\begin{equation}
E_n(q) \ = \ 4\pi \ C_n^2 \ q^{-\alpha} \ \cdot
\label{eq_power_spectr}
\end{equation}
It can then be shown that if $0 < \alpha < 2$,
the phase structure function is similarly a power law of $r$, 
with the opposite power-law index:
\begin{equation}
{\cal S}_\phi(r) \ = \ 
{\rm fc}(\alpha) \ r_{\rm e}^2 \ \lambda^2 \ {\rm SM} \ r^{\alpha} \ ,
\label{eq_struct_power_law}
\end{equation}
where ${\rm fc}(\alpha)$ is a known function of $\alpha$ and
\begin{equation}
{\rm SM} \ \equiv \ \int_{\rm src}^{\rm obs} C_n^2 \ ds
\label{eq_SM}
\end{equation}
is the so-called scattering measure \citep{rickett_90}.
Thus, by measuring the phase structure function at various baseline lengths,
one can in principle retrieve 
both the slope and the line-of-sight--integrated amplitude 
of the electron density power spectrum.

If the power law describing the electron density power spectrum 
is truncated at a minimum wavenumber $q_{\rm min}$
(corresponding to an outer scale
$\ell_{\rm out} \approx \frac{\displaystyle 1}{\displaystyle q_{\rm min}}$)
and a maximum wavenumber $q_{\rm max}$
(corresponding to an inner scale
$\ell_{\rm in} \approx \frac{\displaystyle 1}{\displaystyle q_{\rm max}}$),
Eq.~(\ref{eq_struct_power_law}) remains approximately valid in the range
$\ell_{\rm in} < r < \ell_{\rm out}$,
but ${\cal S}_\phi(r)$ falls off faster to $0$ below $\ell_{\rm in}$ 
and saturates above $\ell_{\rm out}$ \citep{rickett_90,haverkorn&s_13}
(see Figure~\ref{figure_visibility}). 
Hence, the inner and outer scales of the electron density power spectrum 
can in principle be deduced from the low and high values of $r$
at which the phase structure function departs from its power-law behavior.
More precisely, the outer scale can in principle be defined by the value of $r$
at which the phase structure function reaches half its saturated value 
(which is twice the phase variance), i.e.,
${\cal S}_\phi(\ell_{\rm out}) = \langle \delta \phi ^2 \rangle$
\citep{rickett_90}.
In practice, however, only the inner scale is accessible in this manner --
the outer scale is by many orders of magnitude larger 
than the longest interferometer baselines (see below).
Existing measurements indicate that the inner scale of electron density 
turbulence in the ISM is roughly $\sim 100~{\rm km}$.
For instance, \cite{spangler&g_90} and \cite{molnar&mrj_95} obtained 
$\ell_{\rm in} \approx (50-200)~{\rm km}$
and $\ell_{\rm in} \approx 300~{\rm km}$, respectively.
These values of $\ell_{\rm in}$ turn out to be close to those of
the proton inertial length, $l_{\rm p} =
\frac{\displaystyle V_{\rm A}}{\displaystyle \Omega_{\rm p}}$,
and the proton gyro-radius, $r_{\rm p} = 
\frac{\displaystyle \upsilon_{\rm \perp p}}{\displaystyle \Omega_{\rm p}}$, in the WIM,
which, for the parameter values listed in Table~\ref{table_plasma_parameters},
are $l_{\rm p} \simeq 420~{\rm km}$ and $r_{\rm p} \simeq 240~{\rm km}$, respectively.

It is possible to study electron density fluctuations on scales 
(much) larger than the longest interferometer baselines
by analyzing the  intensity ($I$) angular distribution of a scattered radio point source.
Following \cite{rickett_90}, we distinguish between the regimes 
of weak and strong interstellar scintillation (ISS),
based on the value of the scintillation index,
\begin{equation}
m \equiv \frac{\delta I_{\rm rms}}{\langle I \rangle} \ \cdot
\label{eq_scint_index}
\end{equation}

In the weak ISS regime ($m \ll 1$),
the point source shows intensity variations on scales of the order 
of the Fresnel scale,
\begin{equation}
R_{\rm F} \equiv \sqrt{\frac{L}{k}} \ ,
\label{eq_Fresnel}
\end{equation}
where $L$ is the effective propagation path length
through the scattering medium and, as before, 
$k$ is the wavenumber of the incoming radio wave
\citep{rickett_90}.
These intensity variations are produced by electron density fluctuations 
on the same scales, so they provide a direct measure of $\delta n_{\rm e}$ 
around the Fresnel scale.
Typically, for a $1~{\rm GHz}$ radio source 
located at a $1~{\rm kpc}$ effective distance,
$R_{\rm F} \simeq 1.2 \times 10^{6}~{\rm km}$, 
corresponding to an angular scale $\simeq 8~{\rm \mu as}$.

In the strong ISS regime ($m \simeq 1$),
the intensity distribution of the point source has a two-scale pattern,
with the smaller scales $\sim R_{\rm coh}$ representative of diffractive ISS
and the larger scales $\sim R_{\rm sc}$ representative of refractive ISS.
Here, $R_{\rm coh}$ is the electromagnetic field coherence scale, 
defined as the spatial separation (at an observing plane)
across which the r.m.s. phase difference is $1~{\rm rad}$,
i.e., according to Eq.~(\ref{eq_struct_fc}),
${\cal S}_\phi(R_{\rm coh}) = 1~{\rm rad^2}$.
$R_{\rm sc} = L \, \theta_{\rm sc}$ is the radius of the so-called scattering disk,
which represents the sky area around the observed point source 
from which radiation is received, and 
\begin{equation}
\theta_{\rm sc} = \frac{1}{k \, R_{\rm coh}}
\label{eq_scatter_angle}
\end{equation}
is the effective scattering angle.
There is again a relationship between the spectra of $\delta I$ and $\delta n_{\rm e}$, 
but this relationship is much less straightforward than in weak ISS.
Roughly speaking, diffractive and refractive intensity variations 
reflect electron density fluctuations on scales $\sim R_{\rm coh}$ 
and $\sim R_{\rm sc}$, respectively.
Typically, $R_{\rm coh} \sim (10^{3}-10^{5})~{\rm km}$
and $R_{\rm sc} \sim (10^{7}-10^{9})~{\rm km}$
\citep{rickett_88}.

It should be noted that in weak ISS, $R_{\rm coh} > R_{\rm F}$,
while in strong ISS, $R_{\rm coh} < R_{\rm F} < R_{\rm sc}$
and $R_{\rm coh} \, R_{\rm sc} = R_{\rm F}^2$.

\begin{figure}
\includegraphics[height=0.5\textheight]{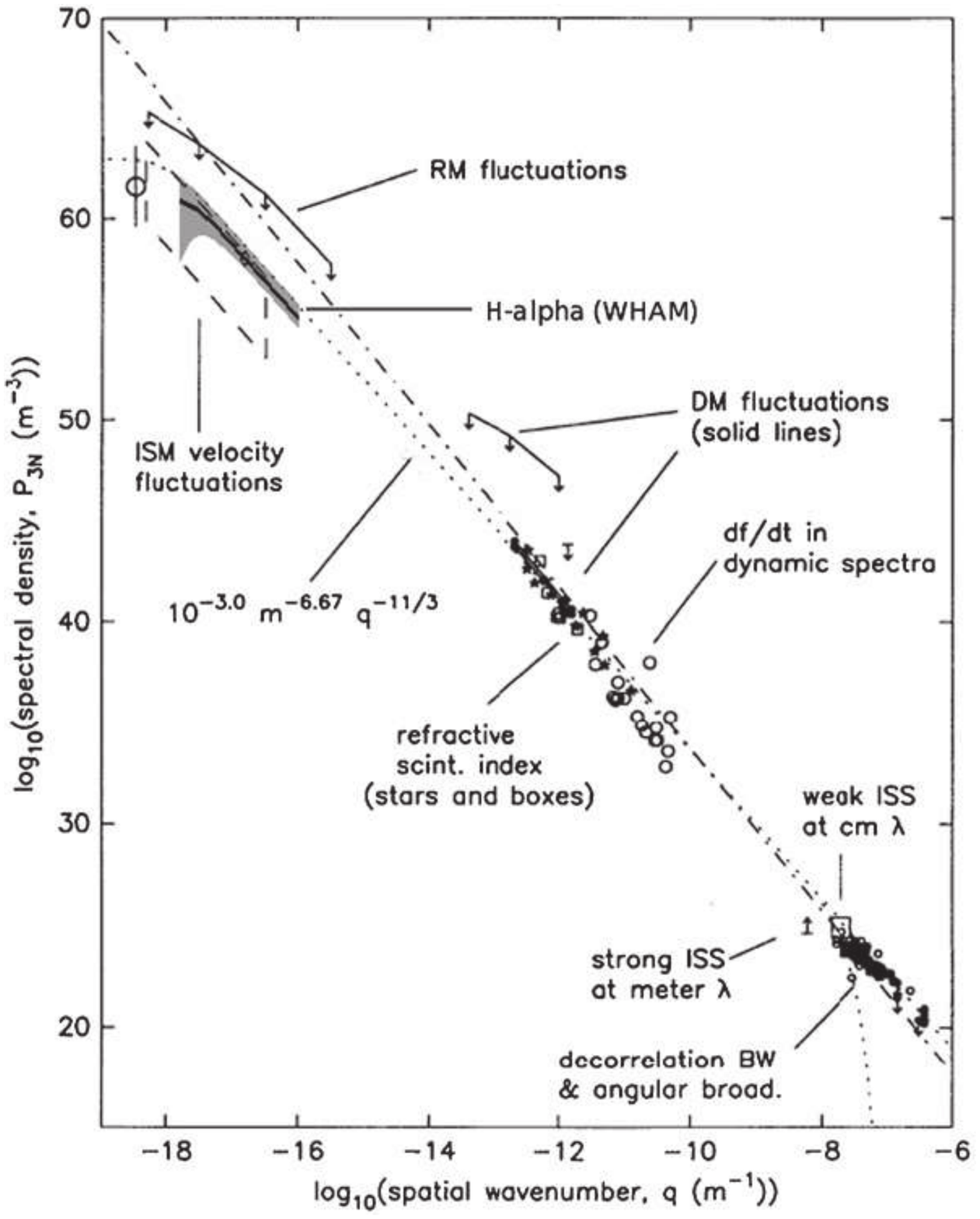}
\caption{Composite spatial power spectrum of electron density fluctuations 
in the nearby ISM, based on a combination of interstellar scattering data 
and measurements at larger scales \citep{armstrong&rs_95,chepurnov&l_10}.
}
\label{figure_big_power_law}
\end{figure}

\cite{armstrong&rs_95} took advantage of the various ISS phenomena 
occurring on widely different scales to construct a composite power spectrum 
of the electron density in the nearby ($\lesssim 1~{\rm kpc}$) ISM,
over scales ranging from $\simeq 2\,000~{\rm km}$ to $\simeq 10^{10}~{\rm km}$ 
($\approx 100~{\rm AU}$).
They found that their composite spectrum was consistent with a single power law, 
$E_n(q) \propto q^{-\alpha}$, with $\alpha \simeq 1.7$, 
over at least the 5 decades from $\simeq 10^5~{\rm km}$ to $\simeq 10^{10}~{\rm km}$,
and most likely down to $\simeq 2\,000~{\rm km}$.
Their derived spectral index is very close to the Kolmogorov value, 
$\alpha = 5/3$.
When combining their ISS data with measurements at larger scales 
(fluctuations in extragalactic-source rotation measures,
gradients in the average electron density), 
they found that their composite power spectrum remained consistent with 
a Kolmogorov-like power law up to $\gtrsim 10^{15}~{\rm km}$ 
($\simeq 30~{\rm pc}$).
This {\it Big power law in the sky}, as the spectrum is now known, 
is displayed in Figure~\ref{figure_big_power_law}.

Later, \cite{chepurnov&l_10} were able to match to the {\it Big power law in the sky}
the spectrum of electron density fluctuations at large scales 
inferred from H$\alpha$ emission data,
thereby confirming that the interstellar Kolmogorov spectrum might extend 
up to $\sim 3~{\rm pc}$.
As we will see in Section~\ref{sect_faraday}, an outer scale to the Kolmogorov
spectrum $\sim 3~{\rm pc}$ is compatible with rotation measure data,
but an outer scale $\sim 30~{\rm pc}$ is not,
as the inferred spectrum of $\delta n_{\rm e}$ appears to flatten out above a few pc.
Such a spectral break could easily be understood if turbulent energy 
is injected into the ISM over a range of scales from a few pc up.

More intriguing is the huge span of the Kolmogorov spectrum,
which suggests that the turbulent cascade proceeds undamped
over more than 10 decades in wavenumber,
despite several well-identified dissipative processes
(reviewed in Appendix~A of \cite{jean&gmf_09}).
In the WIM, the main damping mechanism is viscous damping 
in the collisional regime ($\ell > \lambda_{\rm p}$)
and linear Landau damping in the collisionless regime ($\ell < \lambda_{\rm p}$).
For the parameter values listed in Table~\ref{table_plasma_parameters},
the proton collision mean free-path, $\lambda_{\rm p}$, which separates 
both regimes, is $\simeq 3.5 \times 10^{7}~{\rm km}$ ($\simeq 0.24~{\rm AU}$)
in the WIM.
Now, it is likely that turbulent energy is injected into the ISM 
in the form of the three MHD wave modes.
The fast and slow magnetosonic modes will be mostly dissipated by viscous damping 
at scales $\gg \lambda_{\rm p}$, i.e., before ever reaching the collisionless regime.
On the other hand, the Alfv{\'e}n mode
will cascade almost undamped down to $\lambda_{\rm p}$, 
then enter the collisionless regime,
where it will eventually be dissipated by Landau damping 
at a scale close to the proton inertial length, 
$l_{\rm p} = \frac{\displaystyle V_{\rm A}}{\displaystyle \Omega_{\rm p}}$.
This scenario might explain why the measured inner scale of interstellar turbulence,
$l_{\rm in}$, is found to be close to $l_{\rm p}$ (see paragraph following Eq.~(\ref{eq_SM})).

\subsection{Faraday rotation}
\label{sect_faraday}

Let us now turn to the effects of the interstellar magnetic field
on radio wave propagation. 
To second order (more exactly, to order $1.5^{+}$) in the plasma correction, 
the dispersion relation, Eq.~(\ref{eq_DR}), becomes
\begin{equation}
\omega^2 
\ = \ c^2 \, k^2 + \omega_{\rm e}^2
\mp \frac{\omega_{\rm e}^2 \, \Omega_{\rm e}}{\omega} \ \cdot
\label{eq_DR_2nd}
\end{equation}
The $-$ and $+$ signs in the last term of Eq.~(\ref{eq_DR_2nd}) refer to
the two directions of circular polarization (right and left, respectively).
The reason for the difference between both polarization directions 
is that the electric vector of the right mode rotates in the same direction 
as the electrons gyrate around magnetic field lines,
while the electric vector of the left mode rotates in the opposite direction.
Accordingly, the right and left modes have slightly different phase velocities,
\begin{equation}
V_\phi
\ = \ \frac{\omega}{k} 
\ = \ c \ 
\left( 1 +\frac{\omega_{\rm e}^2}{2 \, \omega^2} 
\mp \frac{\omega_{\rm e}^2 \, \Omega_{\rm e}}{2 \, \omega^3}
\right) \ ,
\label{eq_phase_vel}
\end{equation}
and a phase difference arises between them:
\begin{equation}
\Delta \phi
\ = \ \int_{\rm src}^{\rm obs} \frac{\Delta V_\phi}{c} \ k \ ds
\ = \ \int_{\rm src}^{\rm obs} 
\frac{\omega_{\rm e}^2 \, |\Omega_{\rm e}|}{\omega^3} \ k \ ds \ \cdot
\label{eq_phase_diff}
\end{equation}
Strictly speaking, Eqs.~(\ref{eq_DR_2nd}) -- (\ref{eq_phase_diff}) 
are valid for wave propagation in the direction of the magnetic field, 
$\boldvec{B}$.
For other propagation directions, these equations remain valid 
provided the total magnetic field strength, $B$ (hidden in the expression 
of $\Omega_{\rm e}$) be replaced by the line-of-sight component, $B_\parallel$ 
(counted positively if $\boldvec{B}$ points from the source to the observer).

Now consider a source of linearly polarized radiation, e.g., a Galactic pulsar.
A linearly polarized wave can be regarded as the superposition of a right 
and a left circularly polarized mode.
Since the right mode travels slightly faster than the left mode, 
the direction of linear polarization gradually rotates
as the wave propagates through the plasma.
This effect is known as {\it Faraday rotation}.
The Faraday rotation angle over the entire path from the source to the observer 
is half the phase difference between the right and left modes, 
i.e., in view of Eq.~(\ref{eq_phase_diff}):
\begin{equation}
\Delta \psi 
\ = \ {\rm RM} \
\lambda^2 \ ,
\label{eq_rot_angle}
\end{equation}
where
\begin{equation}
{\rm RM} \ \equiv \ 
\frac{e^3}{2\pi \, m_{\rm e}^2 \, c^4} \
\int_{\rm src}^{\rm obs} n_{\rm e} \ B_\parallel \ ds
\label{eq_RM}
\end{equation}
is the so-called rotation measure.
Thus, the polarization angle of the incoming radiation,
\begin{equation}
\psi_{\rm obs}
\ = \ \psi_{\rm src} + \Delta \psi
\ = \ \psi_{\rm src} + \ {\rm RM} \ \lambda^2 \ ,
\label{eq_pol_angle}
\end{equation}
varies linearly with wavelength squared.
By measuring $\psi_{\rm obs}$ at several wavelengths, 
one can in principle determine both the polarization angle at the source, 
$\psi_{\rm src}$, and the rotation measure of the source, ${\rm RM}$. 
The latter, in turn, provides combined information on the electron density,
$n_{\rm e}$, and the line-of-sight magnetic field, $B_\parallel$,
which is generally not easy to disentangle.
In this respect, Galactic pulsars present the advantage that 
their rotation measure (Eq.~\ref{eq_RM}) 
divided by their dispersion measure (Eq.~\ref{eq_DM}) 
directly yields the average (weighted by $n_e$) value of $B_\parallel$ 
between them and the observer.

Rotation measures of both Galactic pulsars and extragalactic radio point sources
have been used to investigate the properties of plasma turbulence in the ISM.
\cite{minter&s_96} combined the rotation measures of 38 extragalactic sources,
located in a small high-latitude area of the sky, with emission measures,
\begin{equation}
{\rm EM} \ \equiv \ \int n_{\rm e}^2 \ ds \ ,
\label{eq_EM}
\end{equation}
deduced from the observed H$\alpha$ emission in the same area.
This enabled them to derive the structure functions of RM and EM, 
and hence the power spectra of $\delta n_{\rm e}$ and $\delta B$, separately.
They found that the RM and EM structure functions could both be described
by the same broken power law,
\begin{equation}
{\cal S}_{\rm RM}(\delta \theta) , \ {\cal S}_{\rm EM}(\delta \theta)  
\ \propto \ 
\left\lbrace
\begin{array}{l}
\delta \theta^{\, \frac{5}{3}} \qquad {\rm for} \quad \delta \theta < 0.07^\circ \\
\noalign{\smallskip}
\delta \theta^{\, \frac{2}{3}} \qquad {\rm for} \quad \delta \theta > 0.07^\circ 
\end{array}
\right. \ ,
\label{eq_struct_fcs}
\end{equation}
with $\delta \theta$ the angular separation in the sky,
and similarly for the power spectra of $\delta n_{\rm e}$ and $\delta B$:
\begin{equation}
E_n(q) , \ E_B(q) 
\ \propto \ 
\left\lbrace
\begin{array}{l}
q^{\, -\frac{5}{3}} \qquad {\rm for} \quad \ell < 3.6~{\rm pc} \\
\noalign{\smallskip}
q^{\, -\frac{2}{3}} \qquad {\rm for} \quad 3.6~{\rm pc} < \ell \lesssim 100~{\rm pc}
\end{array}
\right. \ ,
\label{eq_power_spectra}
\end{equation}
where, as before, $q \approx \frac{\displaystyle 1}{\displaystyle \ell}$
is the fluctuation wavenumber.
Only the large-scale portion of the spectra was actually fitted 
to the RM and EM data;
the small-scale portion (out of reach of RM studies) was assumed to be Kolmogorov, 
with the amplitude of $E_n(q)$ matched to the {\it Big power law in the sky}.
The authors argued that the spectral break at $\ell \simeq 4 \ {\rm pc}$ 
could reflect a transition from 3D turbulence at small scales 
to 2D turbulence at larger scales, possibly due to the existence 
of turbulent sheets or filaments of thickness $\sim 4 \ {\rm pc}$.
They also showed that their structure functions taken together
could not be reproduced with density fluctuations alone,
but that magnetic field fluctuations were required,
with $\delta B_{\rm rms} \approx 1~\mu{\rm G}$ for the Kolmogorov portion
of the spectrum.
This result indicates that interstellar turbulence is truly MHD in nature.

\cite{haverkorn&bgm08} estimated the outer scale of plasma turbulence in the ISM
based on the structure function of extragalactic-source rotation measures.
They found significant differences between spiral arms and interarm regions.
Both could be ascribed Kolmogorov-like behaviors 
(${\cal S}_{\rm RM}(\delta \theta) \propto \delta \theta^{5/3}$)
up to a $\delta \theta$ corresponding to $\ell \sim {\rm a~few~pc}$, 
but in spiral arms ${\cal S}_{\rm RM}(\delta \theta)$ remained approximately flat 
above $\ell \sim {\rm a~few~pc}$, 
while in interarm regions ${\cal S}_{\rm RM}(\delta \theta)$ kept increasing,
albeit with a shallower slope, up to $\ell \sim 100~{\rm pc}$.
The authors concluded that the outer scale of plasma turbulence is 
$\ell_{\rm out} \sim {\rm a~few~pc}$ in spiral arms 
and $\ell_{\rm out} \sim 100~{\rm pc}$ in interarm regions.
Their interpretation was that the injection of turbulent energy into the ISM
is dominated by stellar winds and protostellar outflows (acting on parsec scales) 
in spiral arms and by supernova and superbubble explosions 
(acting on $\sim 100~{\rm pc}$ scales) in interarm regions.

Information on the magnetic energy spectrum at larger scales comes from 
the rotation measures of Galactic pulsars, combined with their dispersion measures
and estimated distances.
Using a set of 490 pulsars distributed over roughly one third of the Galactic disk, 
\cite{han&fm_04} obtained $E_B(q) \propto q^{-0.37}$, 
with $\delta B_{\rm rms} \approx 6~\mu{\rm G}$,
over the scale range $0.5~{\rm kpc} \lesssim \ell \lesssim 15~{\rm kpc}$.
Since the upper scale is close to the size of the Galaxy, 
this spectrum must include a contribution from the large-scale 
Galactic magnetic field.
The authors suggested that their nearly flat magnetic energy spectrum 
could be the result of an inverse cascade of magnetic helicity 
from the turbulent energy injection scales up to the large Galactic scales.
However, they also remarked that the important amplitude discontinuity 
between their spectrum for $\ell \gtrsim 0.5~{\rm kpc}$ 
and that of \cite{minter&s_96} for $\ell \lesssim 100~{\rm pc}$ 
pointed to a substantial fraction of the energy input to the large-scale 
magnetic field occurring directly at Galactic scales, e.g., 
through the large-scale shear associated with Galactic differential rotation.

Pulsar RMs face the same difficulty as pulsar DMs 
(see Section~\ref{sect_dispersion}):
they are inherently too sparse to give access to the small-scale structure 
of the interstellar magnetic field.
Extragalactic-source RMs fare better in that respect.
There are currently $\approx 45\,000$ extragalactic point sources
with measured RMs \citep{oppermann&jge_15,schnitzeler&cwg_19}.
With SKA\,1, this number is expected to go up to $\sim (1-4) \times 10^7$, 
corresponding to an average surface density $\sim (300 - 1\,000)~{\rm deg}^{-2}$
\citep{haverkorn&etal_15}.
Therefore, the expected angular resolution will be $\sim 2~{\rm arcmin}$,
which, at a typical distance $\sim 1~{\rm kpc}$, implies a spatial resolution 
$\sim 0.6~{\rm pc}$.

\section{Radio polarized emission}
\label{sect_polar}

\subsection{Synchrotron emission}
\label{sect_synchr}

Synchrotron emission is produced by relativistic electrons gyrating
about magnetic field lines.
The synchrotron emissivity at frequency $\nu$ due to a power-law energy
spectrum of relativistic electrons, $f(E) = K_e \, E^{-\gamma}$,
is given by
\begin{equation}
{\cal E}_\nu 
\ = \ 
{\rm fc}(\gamma) \ K_e \ B_\perp^{\gamma + 1 \over 2} \
\nu^{- {\gamma - 1 \over 2}} \ ,
\label{eq_synchr_emiss}
\end{equation}
where ${\rm fc}(\gamma)$ is a known function of $\gamma$
and $B_\perp$ is the strength of the magnetic field projected 
onto the plane of the sky \citep{ginzburg&s_65}.
The synchrotron intensity is obtained by integrating the synchrotron emissivity
along the line of sight:
\begin{equation}
I_\nu 
\ = \ 
\int {\cal E}_\nu \ ds \ \cdot
\label{eq_synchr_intensity}
\end{equation}

Low-frequency radio maps of the Galactic synchrotron emission can be used 
to model the spatial distribution of the interstellar magnetic field,
from the large Galactic scales down to the smallest scales resolved by radio-telescopes.
This modeling requires knowing the relativistic electron spectrum, 
which can be derived either from cosmic-ray propagation models
or from gamma-ray observations.
Alternatively, one sometimes resorts to the double assumption that
(1) relativistic electrons represent a fixed fraction of the cosmic-ray population 
and (2) cosmic rays and magnetic fields are in (energy or pressure) equipartition.
While this assumption can find some rough justification at large scales, 
there is no guarantee that it holds at small scales.
Nevertheless, with the cosmic-ray ion and electron spectra directly measured 
by the {\it Voyager} spacecraft, it can be verified that, 
in the Galactic vicinity of the Sun, cosmic rays and magnetic fields 
are indeed close to (pressure) equipartition, 
with a total magnetic field strength $B \approx 5~\mu{\rm G}$
\citep{ferriere_98, burlaga&n_14, cummings&shl16}.

An important property of synchrotron emission is that it is
linearly polarized perpendicular to $\boldvec{B}_\perp$, 
so that information can also be gained on the orientation of $\boldvec{B}_\perp$.
Evidently, if the observing frequency is low enough to be affected 
by Faraday rotation (see Section~\ref{sect_faraday}),
the received polarized signal must somehow be "de-rotated" 
in order to recover the true field orientation.
In addition, if $\boldvec{B}$ has a fluctuating component,
the contributions from isotropic magnetic fluctuations 
to the polarized intensity cancel out, leaving only the contribution
from the ordered (= mean + anisotropic random) magnetic field,
$\boldvec{B}_{\rm ord}$. 
Thus, while the synchrotron {\it total} intensity, $I_\nu$, 
yields the strength of the {\it total} magnetic field,
the synchrotron {\it polarized} intensity (a complex quantity),
\begin{equation}
P_\nu = Q_\nu + i \, U_\nu \ ,
\label{eq_synchr_PI}
\end{equation}
yields the strength and the orientation of the {\it ordered} magnetic field 
(both in the plane of the sky).
In the large-scale vicinity of the Sun, the ratio of ordered to total 
magnetic field strength turns out to be 
$\frac{\displaystyle B_{\rm ord}}{\displaystyle B} \approx 0.6$ \citep{beck_01}.
Together with $B \approx 5~\mu$G, this ratio implies 
$B_{\rm ord} \approx 3~\mu$G.

Fluctuations in synchrotron total intensity, $I_\nu$,
provide useful diagnostics of the statistical properties
of the underlying magnetized turbulence.
\cite{lazarian&p_12} presented a theoretical description of these fluctuations
and showed that they are anisotropic,
forming filamentary structures aligned with the magnetic field.
\cite{herron&blgm_16} tested and confirmed the theoretical predictions 
of \cite{lazarian&p_12} with three-dimensional (3D) MHD simulations.
They also explored the possibility of retrieving the sonic and Alfv{\'e}nic Mach
numbers from synchrotron intensity maps.
In the process, they brought to light a degeneracy between the Alfv{\'e}nic Mach number 
and the inclination of the mean magnetic field to the line of sight,
and they argued that breaking this degeneracy required observations
of the synchrotron polarized intensity.

\cite{iacobelli&hop_13} 
examined a high-resolution image of the diffuse synchrotron emission 
from the highly polarized Fan region, obtained with the LOw Frequency ARray 
(LOFAR). 
They found that the angular power spectrum of the synchrotron total intensity 
follows a power law of multipole degree, $l$:
\begin{equation}
C_l \propto l^{\, -1.84} \ ,
\label{eq_multipole_obs}
\end{equation}
for $100 \lesssim l \lesssim 1\,300$, 
corresponding to $110' \gtrsim \delta \theta \gtrsim 8'$.
They compared their measured spectrum to a simple model of MHD turbulence 
in the Galaxy proposed by \cite{cho&l_02};
assuming that the synchrotron emissivity has, like the magnetic field, 
a Kolmogorov power spectrum ($P_{\rm syn}(q) \propto q^{-11/3}$),
this model predicts
\begin{equation}
C_l \ \propto \ 
\left\lbrace
\begin{array}{l}
l^{\, -\frac{11}{3}} \qquad {\rm for} \quad l > l_{\rm cr} \\
\noalign{\smallskip}
l^{\, -1} \ \, \qquad {\rm for} \quad l < l_{\rm cr}
\end{array}
\right. \ ,
\label{eq_multipole_mod}
\end{equation}
with $l_{\rm cr} \sim \frac{\displaystyle \pi}{\displaystyle \delta \theta_{\rm cr}}$,
$\delta \theta_{\rm cr} \sim \frac{\displaystyle \ell_{\rm out}}{\displaystyle L}$,
$\ell_{\rm out}$ the size of the largest turbulent cells,
and $L$ the distance to the farthest cells.
The two regimes described by Eq.~(\ref{eq_multipole_mod}) represent 
the extreme cases when two lines of sight separated by $\delta \theta$ traverse 
mostly the same large turbulent cells ($\delta \theta < \delta \theta_{\rm cr}$)
or mostly independent large cells ($\delta \theta > \delta \theta_{\rm cr}$),
respectively.
Since the measured spectral index is closer to the prediction
for $l < l_{\rm cr}$, \cite{iacobelli&hop_13} concluded that 
$l_{\rm cr} \gtrsim 1\,300$, 
and by implication, $\ell_{\rm out} \lesssim 20~{\rm pc}$.
This upper limit to the outer scale of turbulence is consistent with estimates
from RM studies (see Section~\ref{sect_faraday}).
\cite{iacobelli&hop_13} also deduced the ratio of ordered to random 
magnetic field strength from the ratio of mean to r.m.s. synchrotron intensity,
multiplied by $\sqrt{\frac{\displaystyle \ell_{\rm out}}{\displaystyle L}}$~:
$\frac{\displaystyle B_{\rm ord}}{\displaystyle \delta B_{\rm rms}} \lesssim 0.3$.\footnote{
The authors mistakenly wrote 
$\frac{\displaystyle B_{\rm ord}}{\displaystyle \delta B_{\rm rms}} \gtrsim 0.3$,
but their derived value is actually an upper limit 
(Marco Iacobelli \& Marijke Haverkorn, private communication).
}

Fluctuations in synchrotron polarized intensity, $P_\nu$, 
are also potent tracers of the statistical properties of magnetized turbulence,
especially when they are observed at various wavelengths \citep{lazarian&p_16}.
High-resolution images of the Galactic synchrotron polarized intensity
reveal a complex network of filamentary structures,
which generally vary with wavelength,
possess no counterparts in total intensity,
and have therefore been attributed to 
small-scale fluctuations in Faraday rotation.
Particularly striking are the so-called depolarization canals
-- dark lanes believed to result from differential Faraday rotation,
either along the line of sight (when Faraday rotation coexists 
with synchrotron emission) or in the plane of the sky (e.g., at the boundary 
or other strong-gradient layer of a foreground Faraday screen)
\citep{fletcher&s_07}.
Clearly, the structures seen in polarization carry information on magneto-ionic 
turbulence in the medium where Faraday rotation occurs, i.e., in the WIM.
\cite{gaensler&etal_11} showed that one way to extract this information
is to consider the gradient (in the plane of the sky) of the polarized intensity.
They obtained a polarization gradient image of a small area of the Galactic plane,
which they compared to the results of 3D MHD simulations. 
The comparison led them to conclude that interstellar turbulence in the WIM
has a relatively low sonic Mach number, ${\cal M}_{\rm s} \lesssim 2$.
This conclusion was later confirmed by \cite{burkhart&lg_12},
using more sophisticated statistical tools to analyze 
the polarization gradient images.

The anisotropic character of magnetized turbulence opens a new avenue 
to trace the orientation of the interstellar magnetic field,
which can be complementary to the more direct observations
of polarization directions (of synchrotron or dust thermal emission).
In particular, strong Alfv{\'e}nic turbulence takes the form of eddy-like motions 
perpendicular to the local magnetic field, which in turn induce velocity gradients 
perpendicular to the local magnetic field.
Hence the idea of measuring velocity gradients, based on spectroscopic data,
to infer the magnetic field orientation in the plane of the sky
\citep{gonzalez&l_17}.
This technique was recently applied to five star-forming molecular clouds 
in the Gould Belt and shown to yield results similar to those obtained 
from dust polarized emission \citep{hu&etal_19}.

In the same spirit, the work of \cite{lazarian&p_12} indicates that 
gradients of synchrotron total intensity, too,
can serve as tracers of the magnetic field orientation 
in the plane of the sky.
\cite{lazarian&ylc_17} successfully tested the concept
both with synthetic maps from 3D MHD simulations
and with the existing {\it Planck} synchrotron maps.
They also discussed the potential of using synchrotron intensity gradients
(which are unaffected by Faraday rotation) in conjunction with synchrotron
polarization directions to quantify Faraday rotation
or, in the presence of Faraday depolarization, to separate the contributions 
from distant (depolarized) and nearby (non depolarized) regions.

Gradients of synchrotron polarized intensity can be employed in a similar manner
\cite{lazarian&y_18}.
In addition, since Faraday depolarization depends on wavelength,
the line-of-sight depth of the layer 
contributing to the measured polarized intensity
also varies with wavelength.
This makes it in principle possible to map out the line-of-sight distribution,
and hence the full 3D distribution, of $\boldvec{B}_\perp$ --
not only its orientation, as we just explained,
but also its strength, provided a suitable assumption is made
for the relativistic electron density (see beginning of Section~\ref{sect_synchr}).
Here, it is assumed that fluctuations in the relativistic electron density 
are negligible at small scales.
Finally, gradients of $\frac{\displaystyle dP_\nu}{\displaystyle d\lambda^2}$
can give access to $B_\parallel$ 
(see Eqs.~(\ref{eq_rot_angle}) -- (\ref{eq_RM})),
which combined with the above $\boldvec{B}_\perp$,
could enable one to reconstruct the full 3D magnetic field vector.

\subsection{Faraday tomography}
\label{sect_tomo}

An important limitation of the observational methods described in the previous sections
(with the exception of the method based on synchrotron polarization gradients; 
see end of Section~~\ref{sect_synchr})
is that they provide only line-of-sight integrated quantities, 
with no details on how the integrants vary along the line of sight.
For instance, the synchrotron intensity (Eq.~\ref{eq_synchr_intensity}) 
measured in a given direction tells us only about the total amount 
of synchrotron emission produced along the entire line of sight 
through the Galaxy, with no information on the local value of 
${\cal E}_\nu$ (Eq.~\ref{eq_synchr_emiss}).
Similarly, the rotation measure of a given radio source (Eq.~\ref{eq_RM}) 
tells us only about the total amount of Faraday rotation incurred 
along the line of sight between the source and the observer, 
with no information on the local value of $n_e \, B_\parallel$.

A new, powerful and promising, approach to probe the 3D structure 
of the interstellar magnetic field is now being increasingly utilized.
This approach is also based on Faraday rotation, 
but instead of considering the Faraday rotation of the linearly polarized radiation 
from a background radio source (as explained in Section~\ref{sect_faraday}), 
the idea is to exploit the Faraday rotation of the synchrotron radiation 
from the Galaxy itself (discussed in Section~\ref{sect_synchr}).

As a reminder, in the case of a background radio source, 
i.e., when the regions of radio emission and Faraday rotation 
are spatially separated, the Faraday rotation angle, $\Delta \psi$, 
varies linearly with wavelength squared, $\lambda^2$,
and one may define the rotation measure, RM, as being the slope 
of the linear relation between $\Delta \psi$ and $\lambda^2$
(see Eq.~(\ref{eq_rot_angle})).
Hence, RM is a purely observational quantity, 
which can be meaningfully measured only for a background radio source 
and which can then be related to the physical properties ($n_e$ and $B_\parallel$)
of the foreground Faraday-rotating medium through Eq.~(\ref{eq_RM}).

In contrast, when the radio source is the Galaxy itself, 
the regions of radio emission and Faraday rotation are spatially mixed. 
In that case, $\Delta \psi$ no longer varies linearly with $\lambda^2$ 
and the very concept of rotation measure becomes meaningless.
However, one may resort to the more general concept of Faraday depth, 
defined as
\begin{equation}
\label{eq_FD}
\Phi(z) \ \equiv \
\frac{e^3}{2\pi \, m_{\rm e}^2 \, c^4} \
\int _0 ^z n_e \ B_\parallel \ ds \ ,
\end{equation}
where all symbols have the same meaning as in Eq.~(\ref{eq_RM}) 
and $z$ is the line-of-sight distance from the observer
\citep{burn_66,brentjens&d_05}.
$\Phi(z)$ has basically the same formal expression as RM (Eq.~\ref{eq_RM}),\footnote{
The order of the integration limits in Eq.~(\ref{eq_RM}) is irrelevant, 
provided one sticks to the convention that $B_\parallel$ is positive (negative) 
if the magnetic field points toward (away from) the observer.
}
but it differs from RM in the sense that it is a truly physical quantity, 
which can be defined at any point of the ISM, 
independent of any background radio source.
$\Phi(z)$ simply corresponds to the line-of-sight depth, $z$, measured 
in terms of Faraday rotation -- in much the same way as optical depth 
corresponds to line-of-sight depth measured in terms of opacity.

When radio emission and Faraday rotation are mixed along the line of sight,
the complex polarized intensity measured at a given wavelength $\lambda$,
$P(\lambda^2)$, is the superposition of the polarized emission produced 
at every line-of-sight distance $z$, i.e., at every Faraday depth $\Phi$,
and Faraday-rotated by an angle $\Delta \psi = \Phi \ \lambda^2$
(from Eq.~(\ref{eq_rot_angle}) with RM replaced by $\Phi$):
\begin{equation}
\label{eq_PI}
P(\lambda^2) \, = \, \int _{-\infty} ^{+\infty} F(\Phi) \ 
e^{2 i \Phi \lambda^2} \ d\Phi \ ,
\end{equation}
where $F(\Phi)$ is the complex polarized intensity per unit $\Phi$, 
also called complex Faraday dispersion function, at $\Phi$ \citep{burn_66}.
Since the Faraday rotation angle varies with wavelength,
the polarized intensities measured at different wavelengths
correspond to different combinations of all the line-of-sight contributions
and, therefore, provide different pieces of information.
Thus, the idea is to measure the polarized intensity at a large number 
of different wavelengths and to convert its variation with $\lambda^2$ 
into a variation with $\Phi$.
Mathematically, this can be done by Fourier-inverting Eq.~(\ref{eq_PI}): 
\begin{equation}
\label{eq_PI_inverted}
F(\Phi) \, = \, \frac{1}{\pi} \ \int _{-\infty} ^{+\infty} P(\lambda^2) \
e^{-2 i \Phi \lambda^2} \ d\lambda^2 \ \cdot
\end{equation}
Obviously, $P(\lambda^2)$ can be measured only for $\lambda^2 \ge 0$,
so one has to make an assumption for $\lambda^2 < 0$.
For instance, one may assume that $P(\lambda^2)$ is Hermitian,
such that $P(-\lambda^2) = P^\star(\lambda^2)$,
in which case $F(\Phi)$ is strictly real \citep{burn_66,brentjens&d_05}.

\begin{figure}[t!]
{\large (a)} \quad 
\raisebox{-0.33\textheight}{
\includegraphics[width=0.7\textwidth]{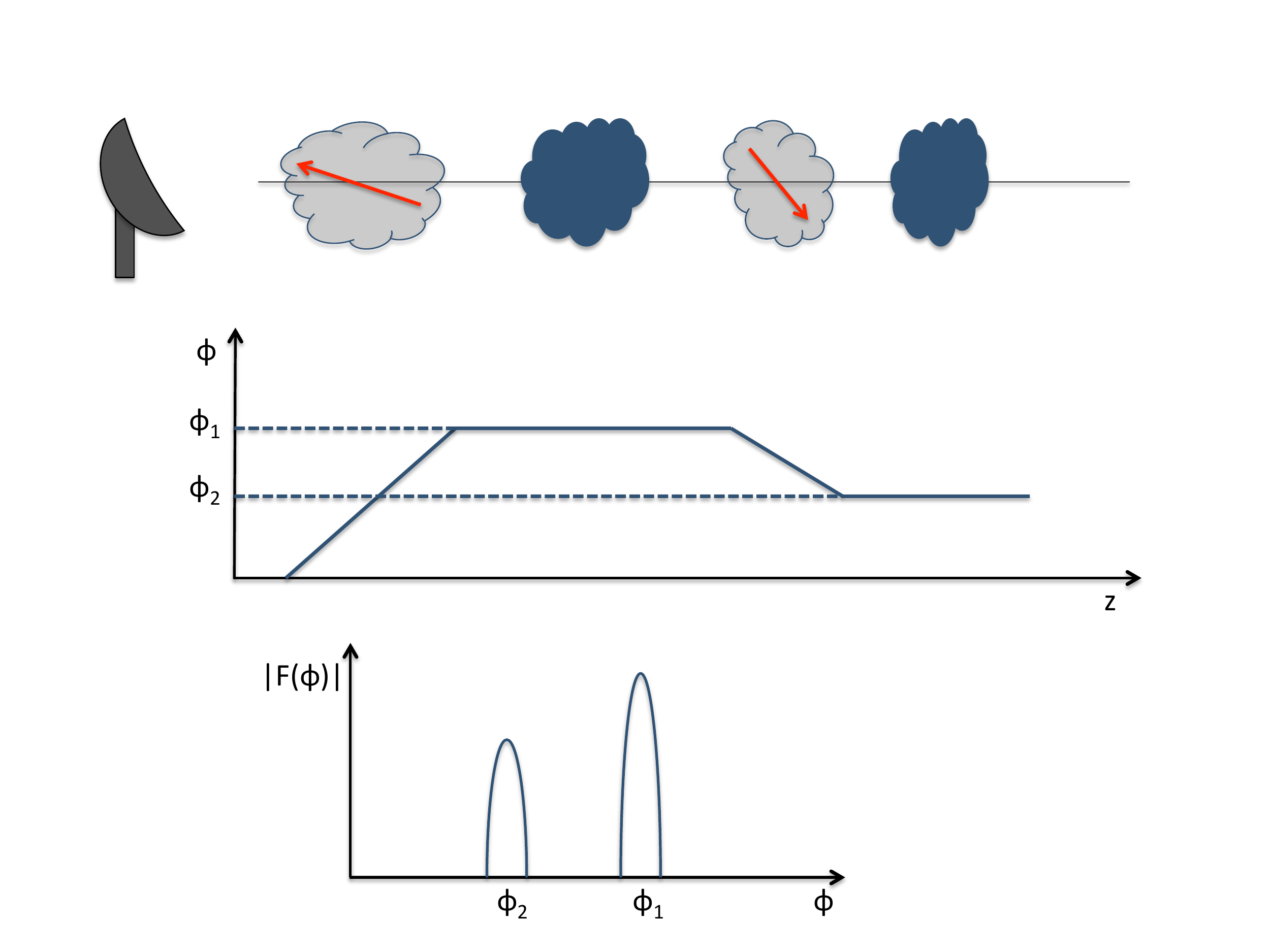}
}
\bigskip \bigskip \bigskip

{\large (b)} \quad 
\raisebox{-0.33\textheight}{
\includegraphics[width=0.7\textwidth]{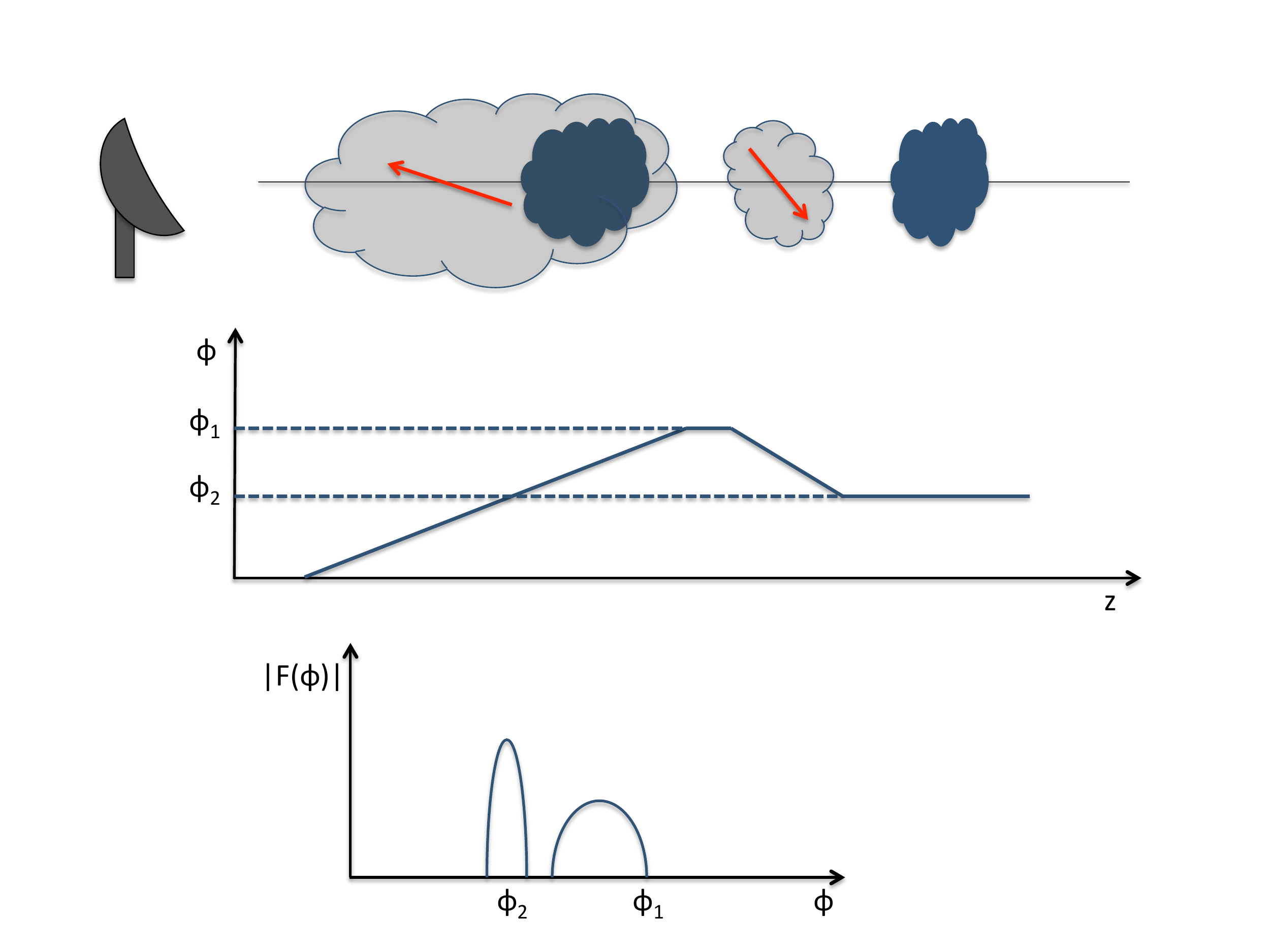}
}
\caption{Schematics illustrating the concept of Faraday tomography.
The top panel in both (a) and (b) pictures the spatial configuration of the system: 
two Faraday-rotating clouds (light grey shading, with a red arrow representing 
the interstellar magnetic field) 
and two synchrotron-emitting clouds (dark blue shading), 
with the observer on the far left.
The middle panel shows how the Faraday depth, $\Phi$ (given by Eq.~(\ref{eq_FD})), 
varies with line-of-sight distance from the observer, $z$.
The bottom panel provides the Faraday dispersion spectrum, $|F(\Phi)|$.
Figure credit: Marta Alves.
}
\label{figure_RMS}
\end{figure}

The method is illustrated in Figure~\ref{figure_RMS}, which depicts a situation 
where the line of sight intersects two Faraday-rotating clouds (shaded in light grey), 
across which $\Phi$ increases or decreases according to Eq.~(\ref{eq_FD}), 
and two synchrotron-emitting clouds (shaded in dark blue).
The top panel in both Figures~\ref{figure_RMS}a and \ref{figure_RMS}b 
indicates the positions of the four clouds along the line of sight 
with respect to the observer (placed on the far left), as well as the directions 
of the magnetic field (red arrows) in the two Faraday-rotating clouds: 
in the closer/farther cloud, the magnetic field points toward/away from the observer, 
so that $B_\parallel$ is positive/negative and $\Phi$ increases/decreases 
with increasing $z$.
The corresponding run of $\Phi$ with $z$ is plotted in the middle panel,
where $\Phi_1$ denotes the Faraday thickness of the closer cloud
and $\Phi_2$ the cumulated Faraday thickness of both Faraday-rotating clouds.
The bottom panel displays the Faraday dispersion spectrum, $|F(\Phi)|$,
with the two peaks representing the polarized emissions 
from the two synchrotron-emitting clouds.
In Figure~\ref{figure_RMS}a, where the Faraday-rotating and synchrotron-emitting
clouds are spatially separated, the closer and farther synchrotron-emitting clouds 
lie at Faraday depths $\Phi_1$ and $\Phi_2$, respectively. 
In Figure~\ref{figure_RMS}b, 
the farther synchrotron-emitting cloud again lies at Faraday depth $\Phi_2$,
but the closer synchrotron-emitting cloud, 
which is now embedded inside a Faraday-rotating cloud, 
extends over a finite range of Faraday depth (up to nearly $\Phi_1$),
i.e., it has a finite Faraday thickness.

In practice, one measures the polarized intensity at many different wavelengths
and Fourier-transforms $P(\lambda^2)$ to obtain $F(\Phi)$.
From the profile of $|F(\Phi)|$, one can then spot synchrotron-emitting regions
and determine their Faraday depths, Faraday thicknesses, and polarized intensities.
One can also uncover intervening Faraday-rotating regions
and determine their Faraday thicknesses.
However, one cannot reconstruct the actual arrangement of the detected 
synchrotron and Faraday regions along the line of sight.
For instance, the profile of $|F(\Phi)|$ in Figure~\ref{figure_RMS}
indicates the presence of two synchrotron regions
at Faraday depths $\Phi_2$ and $\Phi_1$ 
(or up to nearly $\Phi_1$ in Figure~\ref{figure_RMS}b),
as well as the presence of at least two Faraday regions:
one in front of both synchrotron regions and one between them.
However, it does not tell us where the detected regions are located
along the line of sight, how thick they are,
or which of the two synchrotron regions is closer.

Nevertheless, Faraday tomography remains a powerful technique,
especially if the detected synchrotron and Faraday regions
can be identified with known gaseous structures,
because it then offers a new way of tracing their magnetic field.
For synchrotron regions, the derived polarized intensity
can lead to the strength and the orientation of their $\boldvec{B}_\perp$ 
(see Section~\ref{sect_synchr}).
For Faraday regions, the derived Faraday thickness
can lead to their $B_\parallel$ (see Eq.~(\ref{eq_FD})).

Faraday tomography has opened a new window to the mysterious world
of the interstellar magnetic field, furnishing a wealth of details 
on its small-scale structure and its turbulent properties.
Polarization data must be acquired at low radio frequencies in order to enhance 
the effects of Faraday rotation (which increase as $\lambda^2$) 
and, therefore, to improve sensitivity in Faraday space.
At the same time, broad frequency coverage is needed to achieve fine resolution 
in Faraday space \citep{brentjens&d_05}.
This is why low-frequency, broad-band radio-telescopes, 
such as LOFAR \citep{beck&etal_13} and, in the near future, 
SKA\,1 \citep{haverkorn&etal_15}, are ideally suited for the task at hand.

\bibliography{BibTex}

\begin{thebibliography}{56}
\providecommand{\natexlab}[1]{#1}
\providecommand{\url}[1]{\texttt{#1}}
\expandafter\ifx\csname urlstyle\endcsname\relax
  \providecommand{\doi}[1]{doi: #1}\else
  \providecommand{\doi}{doi: \begingroup \urlstyle{rm}\Url}\fi

\bibitem[{Armstrong} et~al.(1995){Armstrong}, {Rickett}, and
  {Spangler}]{armstrong&rs_95}
J.~W. {Armstrong}, B.~J. {Rickett}, and S.~R. {Spangler}.
\newblock {Electron Density Power Spectrum in the Local Interstellar Medium}.
\newblock \emph{\apj}, 443:\penalty0 209, Apr 1995.
\newblock \doi{10.1086/175515}.

\bibitem[{Beck}(2001)]{beck_01}
R.~{Beck}.
\newblock {Galactic and Extragalactic Magnetic Fields}.
\newblock \emph{Space Science Reviews}, 99:\penalty0 243--260, Oct. 2001.

\bibitem[{Beck} et~al.(2013){Beck}, {Anderson}, {Heald}, {Horneffer},
  {Iacobelli}, {K{\"o}hler}, {Mulcahy}, {Pizzo}, {Scaife}, {Wucknitz}, and
  {LOFAR Magnetism Key Science Project Team}]{beck&etal_13}
R.~{Beck}, J.~{Anderson}, G.~{Heald}, A.~{Horneffer}, M.~{Iacobelli},
  J.~{K{\"o}hler}, D.~{Mulcahy}, R.~{Pizzo}, A.~{Scaife}, O.~{Wucknitz}, and
  {LOFAR Magnetism Key Science Project Team}.
\newblock {The LOFAR view of cosmic magnetism}.
\newblock \emph{Astronomische Nachrichten}, 334\penalty0 (6):\penalty0
  548--557, Jun 2013.
\newblock \doi{10.1002/asna.201311894}.

\bibitem[{Brandenburg} and {Lazarian}(2013)]{brandenburg&l_13}
A.~{Brandenburg} and A.~{Lazarian}.
\newblock {Astrophysical Hydromagnetic Turbulence}.
\newblock \emph{\ssr}, 178:\penalty0 163--200, Oct. 2013.
\newblock \doi{10.1007/s11214-013-0009-3}.

\bibitem[{Brentjens} and {de Bruyn}(2005)]{brentjens&d_05}
M.~A. {Brentjens} and A.~G. {de Bruyn}.
\newblock {Faraday rotation measure synthesis}.
\newblock \emph{\aap}, 441:\penalty0 1217--1228, Oct. 2005.
\newblock \doi{10.1051/0004-6361:20052990}.

\bibitem[{Burkhart} et~al.(2012){Burkhart}, {Lazarian}, and
  {Gaensler}]{burkhart&lg_12}
B.~{Burkhart}, A.~{Lazarian}, and B.~M. {Gaensler}.
\newblock {Properties of Interstellar Turbulence from Gradients of Linear
  Polarization Maps}.
\newblock \emph{\apj}, 749\penalty0 (2):\penalty0 145, Apr 2012.
\newblock \doi{10.1088/0004-637X/749/2/145}.

\bibitem[{Burlaga} and {Ness}(2014)]{burlaga&n_14}
L.~F. {Burlaga} and N.~F. {Ness}.
\newblock {Voyager 1 Observations of the Interstellar Magnetic Field and the
  Transition from the Heliosheath}.
\newblock \emph{\apj}, 784:\penalty0 146, Apr. 2014.
\newblock \doi{10.1088/0004-637X/784/2/146}.

\bibitem[{Burn}(1966)]{burn_66}
B.~J. {Burn}.
\newblock {On the depolarization of discrete radio sources by Faraday
  dispersion}.
\newblock \emph{\mnras}, 133:\penalty0 67, 1966.
\newblock \doi{10.1093/mnras/133.1.67}.

\bibitem[{Chepurnov} and {Lazarian}(2010)]{chepurnov&l_10}
A.~{Chepurnov} and A.~{Lazarian}.
\newblock {Extending the Big Power Law in the Sky with Turbulence Spectra from
  Wisconsin H{\ensuremath{\alpha}} Mapper Data}.
\newblock \emph{\apj}, 710\penalty0 (1):\penalty0 853--858, Feb 2010.
\newblock \doi{10.1088/0004-637X/710/1/853}.

\bibitem[{Cho} and {Lazarian}(2002)]{cho&l_02}
J.~{Cho} and A.~{Lazarian}.
\newblock {Magnetohydrodynamic Turbulence as a Foreground for Cosmic Microwave
  Background Studies}.
\newblock \emph{\apjl}, 575\penalty0 (2):\penalty0 L63--L66, Aug 2002.
\newblock \doi{10.1086/342722}.

\bibitem[{Clark} et~al.(2019){Clark}, {Peek}, and
  {Miville-Desch{\^e}nes}]{clark&pm_19}
S.~E. {Clark}, J.~E.~G. {Peek}, and M.~A. {Miville-Desch{\^e}nes}.
\newblock {The Physical Nature of Neutral Hydrogen Intensity Structure}.
\newblock \emph{The Astrophysical Journal}, 874\penalty0 (2):\penalty0 171, Apr
  2019.
\newblock \doi{10.3847/1538-4357/ab0b3b}.

\bibitem[{Cordes} and {Lazio}(2002)]{cordes&l_02}
J.~M. {Cordes} and T.~J.~W. {Lazio}.
\newblock {NE2001.I. A New Model for the Galactic Distribution of Free
  Electrons and its Fluctuations}.
\newblock \emph{ArXiv Astrophysics e-prints}, July 2002.

\bibitem[{Cordes} et~al.(1985){Cordes}, {Weisberg}, and
  {Boriakoff}]{cordes&wb_85}
J.~M. {Cordes}, J.~M. {Weisberg}, and V.~{Boriakoff}.
\newblock {Small-scale electron density turbulence in the interstellar medium}.
\newblock \emph{\apj}, 288:\penalty0 221--247, Jan. 1985.
\newblock \doi{10.1086/162784}.

\bibitem[{Crutcher}(2017)]{crutcher_17}
R.~{Crutcher}.
\newblock {Mapping Magnetic Fields in Molecular Clouds with the CN Zeeman
  Effect}.
\newblock In \emph{72nd International Symposium on Molecular Spectroscopy},
  page TF10, Jun 2017.
\newblock \doi{10.15278/isms.2017.TF10}.

\bibitem[{Crutcher} et~al.(2010){Crutcher}, {Wandelt}, {Heiles}, {Falgarone},
  and {Troland}]{crutcher&whft_10}
R.~M. {Crutcher}, B.~{Wandelt}, C.~{Heiles}, E.~{Falgarone}, and T.~H.
  {Troland}.
\newblock {Magnetic Fields in Interstellar Clouds from Zeeman Observations:
  Inference of Total Field Strengths by Bayesian Analysis}.
\newblock \emph{\apj}, 725\penalty0 (1):\penalty0 466--479, Dec 2010.
\newblock \doi{10.1088/0004-637X/725/1/466}.

\bibitem[{Cummings} et~al.(2016){Cummings}, {Stone}, {Heikkila}, {Lal},
  {Webber}, {J{\'o}hannesson}, {Moskalenko}, {Orlando}, and
  {Porter}]{cummings&shl16}
A.~C. {Cummings}, E.~C. {Stone}, B.~C. {Heikkila}, N.~{Lal}, W.~R. {Webber},
  G.~{J{\'o}hannesson}, I.~V. {Moskalenko}, E.~{Orlando}, and T.~A. {Porter}.
\newblock {Galactic Cosmic Rays in the Local Interstellar Medium: Voyager 1
  Observations and Model Results}.
\newblock \emph{\apj}, 831\penalty0 (1):\penalty0 18, Nov 2016.
\newblock \doi{10.3847/0004-637X/831/1/18}.

\bibitem[{Elmegreen} and {Scalo}(2004)]{elmegreen&s_04}
B.~G. {Elmegreen} and J.~{Scalo}.
\newblock {Interstellar Turbulence I: Observations and Processes}.
\newblock \emph{\araa}, 42\penalty0 (1):\penalty0 211--273, Sep 2004.
\newblock \doi{10.1146/annurev.astro.41.011802.094859}.

\bibitem[{Falgarone} and {Phillips}(1990)]{falgarone&p_90}
E.~{Falgarone} and T.~G. {Phillips}.
\newblock {A Signature of the Intermittency of Interstellar Turbulence: The
  Wings of Molecular Line Profiles}.
\newblock \emph{\apj}, 359:\penalty0 344, Aug 1990.
\newblock \doi{10.1086/169068}.

\bibitem[{Falgarone} et~al.(1991){Falgarone}, {Phillips}, and
  {Walker}]{falgarone&pw_91}
E.~{Falgarone}, T.~G. {Phillips}, and C.~K. {Walker}.
\newblock {The Edges of Molecular Clouds: Fractal Boundaries and Density
  Structure}.
\newblock \emph{\apj}, 378:\penalty0 186, Sep 1991.
\newblock \doi{10.1086/170419}.

\bibitem[{Ferri{\`e}re}(1998)]{ferriere_98}
K.~{Ferri{\`e}re}.
\newblock {Global Model of the Interstellar Medium in Our Galaxy with New
  Constraints on the Hot Gas Component}.
\newblock \emph{\apj}, 497:\penalty0 759--+, Apr. 1998.
\newblock \doi{10.1086/305469}.

\bibitem[{Ferri{\`e}re}(2001)]{ferriere_01}
K.~M. {Ferri{\`e}re}.
\newblock {The interstellar environment of our galaxy}.
\newblock \emph{Reviews of Modern Physics}, 73\penalty0 (4):\penalty0
  1031--1066, Oct 2001.
\newblock \doi{10.1103/RevModPhys.73.1031}.

\bibitem[{Fletcher} and {Shukurov}(2007)]{fletcher&s_07}
A.~{Fletcher} and A.~{Shukurov}.
\newblock {Depolarization canals and interstellar turbulence}.
\newblock In M.~A. {Miville-Desch{\^e}nes} and F.~{Boulanger}, editors,
  \emph{EAS Publications Series}, volume~23 of \emph{EAS Publications Series},
  pages 109--128, Jan 2007.
\newblock \doi{10.1051/eas:2007008}.

\bibitem[{Gaensler} et~al.(2011){Gaensler}, {Haverkorn}, {Burkhart},
  {Newton-McGee}, {Ekers}, {Lazarian}, {McClure-Griffiths}, {Robishaw},
  {Dickey}, and {Green}]{gaensler&etal_11}
B.~M. {Gaensler}, M.~{Haverkorn}, B.~{Burkhart}, K.~J. {Newton-McGee}, R.~D.
  {Ekers}, A.~{Lazarian}, N.~M. {McClure-Griffiths}, T.~{Robishaw}, J.~M.
  {Dickey}, and A.~J. {Green}.
\newblock {Low-Mach-number turbulence in interstellar gas revealed by radio
  polarization gradients}.
\newblock \emph{Nature}, 478\penalty0 (7368):\penalty0 214--217, Oct 2011.
\newblock \doi{10.1038/nature10446}.

\bibitem[{Ginzburg} and {Syrovatskii}(1965)]{ginzburg&s_65}
V.~L. {Ginzburg} and S.~I. {Syrovatskii}.
\newblock {Cosmic Magnetobremsstrahlung (synchrotron Radiation)}.
\newblock \emph{\araa}, 3:\penalty0 297, Jan 1965.
\newblock \doi{10.1146/annurev.aa.03.090165.001501}.

\bibitem[{Gonz{\'a}lez-Casanova} and {Lazarian}(2017)]{gonzalez&l_17}
D.~F. {Gonz{\'a}lez-Casanova} and A.~{Lazarian}.
\newblock {Velocity Gradients as a Tracer for Magnetic Fields}.
\newblock \emph{\apj}, 835\penalty0 (1):\penalty0 41, Jan 2017.
\newblock \doi{10.3847/1538-4357/835/1/41}.

\bibitem[{Han} et~al.(2004){Han}, {Ferriere}, and {Manchester}]{han&fm_04}
J.~L. {Han}, K.~{Ferriere}, and R.~N. {Manchester}.
\newblock {The Spatial Energy Spectrum of Magnetic Fields in Our Galaxy}.
\newblock \emph{\apj}, 610\penalty0 (2):\penalty0 820--826, Aug 2004.
\newblock \doi{10.1086/421760}.

\bibitem[{Haverkorn} and {Spangler}(2013)]{haverkorn&s_13}
M.~{Haverkorn} and S.~R. {Spangler}.
\newblock {Plasma Diagnostics of the Interstellar Medium with Radio Astronomy}.
\newblock \emph{\ssr}, 178\penalty0 (2-4):\penalty0 483--511, Oct 2013.
\newblock \doi{10.1007/s11214-013-0014-6}.

\bibitem[{Haverkorn} et~al.(2008){Haverkorn}, {Brown}, {Gaensler}, and
  {McClure-Griffiths}]{haverkorn&bgm08}
M.~{Haverkorn}, J.~C. {Brown}, B.~M. {Gaensler}, and N.~M. {McClure-Griffiths}.
\newblock {The Outer Scale of Turbulence in the Magnetoionized Galactic
  Interstellar Medium}.
\newblock \emph{\apj}, 680\penalty0 (1):\penalty0 362--370, Jun 2008.
\newblock \doi{10.1086/587165}.

\bibitem[{Haverkorn} et~al.(2015){Haverkorn}, {Akahori}, {Carretti},
  {Ferri{\`e}re}, {Frick}, {Gaensler}, {Heald}, {Johnston-Hollitt}, {Jones},
  {Landecker}, {Mao}, {Noutsos}, {Oppermann}, {Reich}, {Robishaw}, {Scaife},
  {Schnitzeler}, {Stepanov}, {Sun}, and {Taylor}]{haverkorn&etal_15}
M.~{Haverkorn}, T.~{Akahori}, E.~{Carretti}, K.~{Ferri{\`e}re}, P.~{Frick},
  B.~{Gaensler}, G.~{Heald}, M.~{Johnston-Hollitt}, D.~{Jones}, T.~{Landecker},
  S.~A. {Mao}, A.~{Noutsos}, N.~{Oppermann}, W.~{Reich}, T.~{Robishaw},
  A.~{Scaife}, D.~{Schnitzeler}, R.~{Stepanov}, X.~{Sun}, and R.~{Taylor}.
\newblock {Measuring magnetism in the Milky Way with the Square Kilometre
  Array}.
\newblock In \emph{Advancing Astrophysics with the Square Kilometre Array
  (AASKA14)}, page~96, Apr 2015.

\bibitem[{Hennebelle} and {Falgarone}(2012)]{hennebelle&f_12}
P.~{Hennebelle} and E.~{Falgarone}.
\newblock {Turbulent molecular clouds}.
\newblock \emph{Astronomy and Astrophysics Review}, 20:\penalty0 55, Nov 2012.
\newblock \doi{10.1007/s00159-012-0055-y}.

\bibitem[{Herron} et~al.(2016){Herron}, {Burkhart}, {Lazarian}, {Gaensler}, and
  {McClure-Griffiths}]{herron&blgm_16}
C.~A. {Herron}, B.~{Burkhart}, A.~{Lazarian}, B.~M. {Gaensler}, and N.~M.
  {McClure-Griffiths}.
\newblock {Radio Synchrotron Fluctuation Statistics as a Probe of Magnetized
  Interstellar Turbulence}.
\newblock \emph{The Astrophysical Journal}, 822\penalty0 (1):\penalty0 13, May
  2016.
\newblock \doi{10.3847/0004-637X/822/1/13}.

\bibitem[{Hewish} et~al.(1968){Hewish}, {Bell}, {Pilkington}, {Scott}, and
  {Collins}]{hewish&bps_68}
A.~{Hewish}, S.~J. {Bell}, J.~D.~H. {Pilkington}, P.~F. {Scott}, and R.~A.
  {Collins}.
\newblock {Observation of a Rapidly Pulsating Radio Source}.
\newblock \emph{\nat}, 217:\penalty0 709--713, Feb. 1968.
\newblock \doi{10.1038/217709a0}.

\bibitem[{Hu} et~al.(2019){Hu}, {Yuen}, {Lazarian}, {Ho}, {Benjamin}, {Hill},
  {Lockman}, {Goldsmith}, and {Lazarian}]{hu&etal_19}
Y.~{Hu}, K.~H. {Yuen}, V.~{Lazarian}, K.~W. {Ho}, R.~A. {Benjamin}, A.~S.
  {Hill}, F.~J. {Lockman}, P.~F. {Goldsmith}, and A.~{Lazarian}.
\newblock {Magnetic field morphology in interstellar clouds with the velocity
  gradient technique}.
\newblock \emph{Nature Astronomy}, 3:\penalty0 776--782, Jun 2019.
\newblock \doi{10.1038/s41550-019-0769-0}.

\bibitem[{Iacobelli} et~al.(2013){Iacobelli}, {Haverkorn}, {Orr{\'u}}, {Pizzo},
  {Anderson}, {Beck}, {Bell}, {Bonafede}, {Chyzy}, and
  {Dettmar}]{iacobelli&hop_13}
M.~{Iacobelli}, M.~{Haverkorn}, E.~{Orr{\'u}}, R.~F. {Pizzo}, J.~{Anderson},
  R.~{Beck}, M.~R. {Bell}, A.~{Bonafede}, K.~{Chyzy}, and R.~J. {Dettmar}.
\newblock {Studying Galactic interstellar turbulence through fluctuations in
  synchrotron emission. First LOFAR Galactic foreground detection}.
\newblock \emph{\aap}, 558:\penalty0 A72, Oct 2013.
\newblock \doi{10.1051/0004-6361/201322013}.

\bibitem[{Jean} et~al.(2009){Jean}, {Gillard}, {Marcowith}, and
  {Ferri{\`e}re}]{jean&gmf_09}
P.~{Jean}, W.~{Gillard}, A.~{Marcowith}, and K.~{Ferri{\`e}re}.
\newblock {Positron transport in the interstellar medium}.
\newblock \emph{\aap}, 508\penalty0 (3):\penalty0 1099--1116, Dec 2009.
\newblock \doi{10.1051/0004-6361/200809830}.

\bibitem[{Keane} et~al.(2015){Keane}, {Bhattacharyya}, {Kramer}, {Stappers},
  {Keane}, {Bhattacharyya}, {Kramer}, {Stappers}, {Bates}, {Burgay},
  {Chatterjee}, {Champion}, {Eatough}, {Hessels}, {Janssen}, {Lee}, {van
  Leeuwen}, {Margueron}, {Oertel}, {Possenti}, {Ransom}, {Theureau}, and
  {Torne}]{keane&etal_15}
E.~{Keane}, B.~{Bhattacharyya}, M.~{Kramer}, B.~{Stappers}, E.~F. {Keane},
  B.~{Bhattacharyya}, M.~{Kramer}, B.~W. {Stappers}, S.~D. {Bates},
  M.~{Burgay}, S.~{Chatterjee}, D.~J. {Champion}, R.~P. {Eatough}, J.~W.~T.
  {Hessels}, G.~{Janssen}, K.~J. {Lee}, J.~{van Leeuwen}, J.~{Margueron},
  M.~{Oertel}, A.~{Possenti}, S.~{Ransom}, G.~{Theureau}, and P.~{Torne}.
\newblock {A Cosmic Census of Radio Pulsars with the SKA}.
\newblock \emph{Advancing Astrophysics with the Square Kilometre Array
  (AASKA14)}, art.~40, Apr. 2015.

\bibitem[{Lazarian} and {Pogosyan}(2000)]{lazarian&p_00}
A.~{Lazarian} and D.~{Pogosyan}.
\newblock {Velocity Modification of H I Power Spectrum}.
\newblock \emph{\apj}, 537\penalty0 (2):\penalty0 720--748, Jul 2000.
\newblock \doi{10.1086/309040}.

\bibitem[{Lazarian} and {Pogosyan}(2006)]{lazarian&p_06}
A.~{Lazarian} and D.~{Pogosyan}.
\newblock {Studying Turbulence Using Doppler-broadened Lines: Velocity
  Coordinate Spectrum}.
\newblock \emph{\apj}, 652\penalty0 (2):\penalty0 1348--1365, Dec 2006.
\newblock \doi{10.1086/508012}.

\bibitem[{Lazarian} and {Pogosyan}(2012)]{lazarian&p_12}
A.~{Lazarian} and D.~{Pogosyan}.
\newblock {Statistical Description of Synchrotron Intensity Fluctuations:
  Studies of Astrophysical Magnetic Turbulence}.
\newblock \emph{\apj}, 747\penalty0 (1):\penalty0 5, Mar 2012.
\newblock \doi{10.1088/0004-637X/747/1/5}.

\bibitem[{Lazarian} and {Pogosyan}(2016)]{lazarian&p_16}
A.~{Lazarian} and D.~{Pogosyan}.
\newblock {Spectrum and Anisotropy of Turbulence from Multi-frequency
  Measurement of Synchrotron Polarization}.
\newblock \emph{\apj}, 818\penalty0 (2):\penalty0 178, Feb 2016.
\newblock \doi{10.3847/0004-637X/818/2/178}.

\bibitem[{Lazarian} and {Yuen}(2018)]{lazarian&y_18}
A.~{Lazarian} and K.~H. {Yuen}.
\newblock {Gradients of Synchrotron Polarization: Tracing 3D Distribution of
  Magnetic Fields}.
\newblock \emph{\apj}, 865\penalty0 (1):\penalty0 59, Sep 2018.
\newblock \doi{10.3847/1538-4357/aad3ca}.

\bibitem[{Lazarian} et~al.(2017){Lazarian}, {Yuen}, {Lee}, and
  {Cho}]{lazarian&ylc_17}
A.~{Lazarian}, K.~H. {Yuen}, H.~{Lee}, and J.~{Cho}.
\newblock {Synchrotron Intensity Gradients as Tracers of Interstellar Magnetic
  Fields}.
\newblock \emph{The Astrophysical Journal}, 842\penalty0 (1):\penalty0 30, Jun
  2017.
\newblock \doi{10.3847/1538-4357/aa74c6}.

\bibitem[{Manchester} et~al.(2005){Manchester}, {Hobbs}, {Teoh}, and
  {Hobbs}]{manchester&hth_05}
R.~N. {Manchester}, G.~B. {Hobbs}, A.~{Teoh}, and M.~{Hobbs}.
\newblock {VizieR Online Data Catalog: ATNF Pulsar Catalog (Manchester+,
  2005)}.
\newblock \emph{VizieR Online Data Catalog}, 7245, Aug. 2005.

\bibitem[{Minter} and {Spangler}(1996)]{minter&s_96}
A.~H. {Minter} and S.~R. {Spangler}.
\newblock {Observation of Turbulent Fluctuations in the Interstellar Plasma
  Density and Magnetic Field on Spatial Scales of 0.01 to 100 Parsecs}.
\newblock \emph{\apj}, 458:\penalty0 194, Feb 1996.
\newblock \doi{10.1086/176803}.

\bibitem[{Miville-Desch{\^e}nes} et~al.(2003){Miville-Desch{\^e}nes}, {Joncas},
  {Falgarone}, and {Boulanger}]{miville&jfb_03}
M.~A. {Miville-Desch{\^e}nes}, G.~{Joncas}, E.~{Falgarone}, and F.~{Boulanger}.
\newblock {High resolution 21 cm mapping of the Ursa Major Galactic cirrus:
  Power spectra of the high-latitude H I gas}.
\newblock \emph{\aap}, 411:\penalty0 109--121, Nov 2003.
\newblock \doi{10.1051/0004-6361:20031297}.

\bibitem[{Molnar} et~al.(1995){Molnar}, {Mutel}, {Reid}, and
  {Johnston}]{molnar&mrj_95}
L.~A. {Molnar}, R.~L. {Mutel}, M.~J. {Reid}, and K.~J. {Johnston}.
\newblock {Interstellar Scattering toward Cygnus X-3: Measurements of
  Anisotrophy and of the Inner Scale}.
\newblock \emph{\apj}, 438:\penalty0 708, Jan 1995.
\newblock \doi{10.1086/175115}.

\bibitem[{M{\"u}nch}(1958)]{munch_58}
G.~{M{\"u}nch}.
\newblock {Internal Motions in the Orion Nebula}.
\newblock \emph{Reviews of Modern Physics}, 30\penalty0 (3):\penalty0
  1035--1041, Jul 1958.
\newblock \doi{10.1103/RevModPhys.30.1035}.

\bibitem[{Oppermann} et~al.(2015){Oppermann}, {Junklewitz}, {Greiner},
  {En{\ss}lin}, {Akahori}, {Carretti}, {Gaensler}, {Goobar}, {Harvey-Smith},
  and {Johnston-Hollitt}]{oppermann&jge_15}
N.~{Oppermann}, H.~{Junklewitz}, M.~{Greiner}, T.~A. {En{\ss}lin},
  T.~{Akahori}, E.~{Carretti}, B.~M. {Gaensler}, A.~{Goobar},
  L.~{Harvey-Smith}, and M.~{Johnston-Hollitt}.
\newblock {Estimating extragalactic Faraday rotation}.
\newblock \emph{\aap}, 575:\penalty0 A118, Mar 2015.
\newblock \doi{10.1051/0004-6361/201423995}.

\bibitem[{Planck Collaboration}(2018)]{planck19}
{Planck Collaboration}.
\newblock {Planck 2018 results. XII. Galactic astrophysics using polarized dust
  emission}.
\newblock \emph{arXiv e-prints}, art. arXiv:1807.06212, Jul 2018.

\bibitem[{Rickett}(1988)]{rickett_88}
B.~J. {Rickett}.
\newblock {Introduction to the observables in interstellar radiowave
  propagation}.
\newblock In J.~M. {Cordes}, B.~J. {Rickett}, and D.~C. {Backer}, editors,
  \emph{Radio Wave Scattering in the Interstellar Medium}, volume 174 of
  \emph{American Institute of Physics Conference Series}, pages 2--16, Jan
  1988.
\newblock \doi{10.1063/1.37597}.

\bibitem[{Rickett}(1990)]{rickett_90}
B.~J. {Rickett}.
\newblock {Radio propagation through the turbulent interstellar plasma}.
\newblock \emph{\araa}, 28:\penalty0 561--605, 1990.
\newblock \doi{10.1146/annurev.aa.28.090190.003021}.

\bibitem[{Schnitzeler}(2012)]{schnitzeler_12}
D.~H.~F.~M. {Schnitzeler}.
\newblock {Modelling the Galactic distribution of free electrons}.
\newblock \emph{\mnras}, 427:\penalty0 664--678, Nov. 2012.
\newblock \doi{10.1111/j.1365-2966.2012.21869.x}.

\bibitem[{Schnitzeler} et~al.(2019){Schnitzeler}, {Carretti}, {Wieringa},
  {Gaensler}, {Haverkorn}, and {Poppi}]{schnitzeler&cwg_19}
D.~H.~F.~M. {Schnitzeler}, E.~{Carretti}, M.~H. {Wieringa}, B.~M. {Gaensler},
  M.~{Haverkorn}, and S.~{Poppi}.
\newblock {S-PASS/ATCA: a window on the magnetic universe in the Southern
  hemisphere}.
\newblock \emph{\mnras}, 485\penalty0 (1):\penalty0 1293--1309, May 2019.
\newblock \doi{10.1093/mnras/stz092}.

\bibitem[{Spangler} and {Gwinn}(1990)]{spangler&g_90}
S.~R. {Spangler} and C.~R. {Gwinn}.
\newblock {Evidence for an Inner Scale to the Density Turbulence in the
  Interstellar Medium}.
\newblock \emph{\apj}, 353:\penalty0 L29, Apr 1990.
\newblock \doi{10.1086/185700}.

\bibitem[{Thompson} et~al.(1986){Thompson}, {Moran}, and
  {Swenson}]{thompson&ms_86}
A.~R. {Thompson}, J.~M. {Moran}, and G.~W. {Swenson}.
\newblock \emph{{Interferometry and synthesis in radio astronomy}}.
\newblock Wiley-Interscience Publication, New York: Wiley, 1986.

\bibitem[{Yao} et~al.(2017){Yao}, {Manchester}, and {Wang}]{yao&mw_17}
J.~M. {Yao}, R.~N. {Manchester}, and N.~{Wang}.
\newblock {A New Electron-density Model for Estimation of Pulsar and FRB
  Distances}.
\newblock \emph{\apj}, 835:\penalty0 29, Jan. 2017.
\newblock \doi{10.3847/1538-4357/835/1/29}.

\end{thebibliography}

\end{document}